\documentclass[twoside,a4paper]{article}

\usepackage{graphicx}

\usepackage[T2A]{fontenc}
\usepackage[cp866nav]{inputenc}

\usepackage{cmpj2e}

\usepackage{amsmath}
\usepackage{amssymb}
\usepackage{wrapfig}

\providecommand{\abs}[1]{\lvert#1\rvert}

\allowdisplaybreaks[1]

\title[Phase separation in Li$_x$TiO$_2$: A theory]%
{Phase separation in lithium intercalated anatase:\\ A theory}
\author{O.V. Velychko, I.V. Stasyuk}
\address{Institute for Condensed Matter Physics
of the National Academy of Sciences of Ukraine, 1~Svientsitskii
Str., 79011 Lviv, Ukraine}

\begin{document}

\maketitle

\begin{abstract}
Lithium intercalated anatase used in Li-ion batteries has some
special features: coexistence of Li-rich and Li-poor phases as
well as two possible positions for Li ions in the oxygen
tetrahedron. A theoretical description of the compound considering
those peculiarities is presented. As shown by the performed
symmetry analysis, the intercalation induced lattice deformation
can be accompanied by the ordering of antiferroelectric type
(internal piezoeffect). In the following step, a qualitative
illustration of the phase separation in the lithiated anatase is
given within the Landau expansion at the proper choice of
coefficients. A microscopic model for description of the compound
is also proposed which combines features of the Mitsui and
Blume-Emery-Griffits models and utilizes the symmetry analysis
results. Various ground state and temperature-dependent phase
diagrams of the model are studied to find a set of model
parameters corresponding to the lithiated anatase. A phase
separation into the empty and half-filled phases in a wide
temperature range has been found closely resembling the phase
coexistence in the intercalated crystal. In the framework of the
model, the two-position Li subsystem could have the ordering of
ferro- or antiferroelectric types which, however, has not been yet
observed by the experiment.
\pacs 71.20.Tx, 64.60.Cn, 64.60.De
\end{abstract}

\section{Titania in a nutshell}

\subsection{Titanium dioxide: polymorphs, properties and applications}

There are eleven known polymorphs of titanium dioxide (titania).
The most common natural forms (rutile, anatase and brookite) are
just different space arrangements of the TiO$_6$ group where a
titanium cation is located in the centre of the slightly deformed
octahedron shaped by six oxygen anions. Each polymorph has its own
deviations from the ideal octahedron but it is always elongated
along the certain axis. Thus, two (apical) oxygens are further
away from the titanium than four others (equatorial).

Being the most stable polymorphs, rutile and anatase are widely
used and intensively studied. They are very similar in many
details (e.g. the arrangement of atoms and average lengths of
bonds) \cite{tdio:Cangiani04}. However, anatase is 10\% less dense
than rutile and an additional volume is condensed in the voids
affecting the cell averaged properties such as compressibility and
dielectric constant. Moreover, this minor difference becomes
crucial at crystal intercalation.
In both polymorphs, an elementary cell consists of two formula
units but unlike the tetragonal rutile ($P4_2/mnm$) the standard
crystallographic cell of body-centered tetragonal anatase
($I4_1/amd$) is chosen to consist of two elementary cells.

The titanium atoms, and hence, the octahedra, are arranged in such
a way that each oxygen is at the same time an equatorial atom for
one titanium, and an apical one for the other titanium atom in the
same unit cell. Neighboring octahedra are sharing edges and
corners with each other. Two and four edges of each octahedron are
shared in rutile and anatase, respectively. The basic octahedra
are distorted in such a way that each shared edge is shortened,
the other edges being correspondingly elongated. The shortened
oxygen-oxygen bonds are  often named as the bridge bonds (in the
sense that it bridges the interaction between Ti ions:
metal-oxygen-metal).
In rutile the bridge bond connects two equatorial oxygen atoms.
Hence, the octahedra form the vertical linear chains. The
octahedra belonging to adjacent chains are connected only through
one corner: an oxygen atom which is both apical and equatorial for
the two touching octahedra.
In anatase, the octahedra are arranged in order to share a
diagonal edge between an apical and an equatorial atom. Thus,
octahedra form zig-zag chains orthogonal to the crystallographic
axis. There are two sets of chains orthogonal to each other, that
are connected through a common octahedron.

The list of hi-tech applications of titanium dioxide is quite
impressive. They are primarily related to its photoactivity. For
example, TiO$_2$, particularly in the anatase form, is a
photocatalyst under ultraviolet light. The strong oxidative
potential of the positive holes oxidizes water to create hydroxyl
radicals (the Honda-Fujishima effect \cite{tdio:Fujishima72}).

Superhydrophilicity phenomenon for glass coated with titanium
dioxide is caused by ultraviolet light partially removing oxygen
atoms from the surface of the titanium oxide. The areas where
oxygen atoms were removed became hydrophilic, while the same size
areas where no oxygen atoms were taken away turned out to be
hydrophobic \cite{tdio:Wang97}. The result is a TiO$_2$-coated
glass which is antifogging and self-cleaning.

Dielectric properties of titanium dioxide distinguish it as
semiconductor \cite{tdio:Earle42} (to stress the difference:
anatase is semiconductor of the n-type while rutile is of p-type
which is utilized in the gas sensor \cite{tdio:Savage01}). Due to
its high dielectric constant, it is commonly used as a dielectric
in electronic devices, such as thin film capacitors
\cite{tdio:Wu90} and MOS devices \cite{tdio:Brown78}, as well as
for the fabrication of anti-reflection coatings, interference
filters \cite{tdio:Badwey91}, as well as optical wave-guides
\cite{tdio:Siefering90}.

Although both rutile and anatase are potentially interesting for
photo-catalysis and photo-electrochemical applications,
experimental investigations have mostly focused on the more
prospective anatase polymorph. It has a wider optical-absorption
gap and a smaller electron effective mass which presumably leads
to a higher mobility for the charge carriers \cite{tdio:Tang1994}
and plays a key role in the injection process of novel
dye-sensitized photochemical solar cells with high conversion
efficiency \cite{tdio:Gratzel91}.
In some cases, the two materials are used together in the same
device, exploiting their peculiar properties for different
purposes. For example, a typical low-cost photo-voltaic module is
composed of a transparent conducting photo-electrode of
dye-sensitized nanocrystalline anatase, a spacer of electrically
insulating, light-reflecting particles of rutile, and a
counter-electrode of graphite powder \cite{tdio:Kay96}. Hence,
anatase is used due to its efficient coupling with the dye, and
rutile for its high dielectric constant.

Furthermore, the open crystallographic structure of anatase
facilitates the accommodation of substantial amounts of small ions
(Li, H, etc.) within the lattice. Lithium insertion changes the
optical properties of TiO$_2$: it turns the white powder dark blue
whilst in thin film form it changes from being transparent to
partially reflecting (electrochromism) which is used in displays
and sun-blinds (switchable mirrors). Combining an electrochromic
film and a photovoltaic film to form the two electrodes of an
electrochemical cell one can achieve a photochromic structure
\cite{tdio:Bechinger96}.

\subsection{Intercalation of rutile and anatase: experiment and theory}

In the last decades the Li-ion batteries have run into operation
as a result of their high energy capacity, re-chargeability and
environmentally friendly properties. Anatase TiO$_2$ may act as an
anode in such a battery \cite{tdio:Huang95}.
In practice, anatase is not the ideal candidate because of its
relatively low potential versus other electrode materials. Better
properties are demonstrated, e.g.\ by similar compounds with the
spinel structure LiTi$_2$O$_4$ \cite{tdio:Wagemaker05},
Li$_4$Ti$_5$O$_{12}$ \cite{tdio:Rho06} and their manganesian
analogues LiMn$_2$O$_4$ and LiMg$_{0.1}$Ni$_{0.4}$Mn$_{1.5}$O$_4$
\cite{tdio:Wagemaker04}.
Here anatase is considered as a well defined model material
displaying many typical properties of transition metal oxide
electrodes. The electrode and electrochromic properties of
lithiated anatase are already well documented and partly exploited
commercially. However, the impressive experimental breakthrough in
the study of the microscopic processes resulting in these
achievements is not accompanied by theoretical investigations.

As mentioned earlier, anatase TiO$_2$ has a body-centered
tetragonal structure indexed by the $I4_1/amd$ space group. Upon
lithiation, anatase lattice undergoes an orthorhombic distortion
that results in the Li$_{0.5}$TiO$_{2}$ phase (sometimes referred
to as Li-titanate) indexed by the space group $Imma$, where the
fourth order axis is lost due to the distortion in the $ab$ plane
\cite{tdio:Cava84}. The change in symmetry is accompanied by a
decrease of the unit cell along the $c$-axis and by an increase
along the $b$-axis, resulting in a 4\% increase of the unit cell
volume. Lithium was found to reside in the interstitial voids
within the oxygen octahedra \cite{tdio:Cava84}. The structural
change can be explained as occupation of Ti-Ti bonding atomic
orbitals by the electron that enters the TiO$_2$ lattice with each
Li-ion to maintain charge neutrality.

In comparison with the number of experimental studies of
intercalated titanium dioxide the list of theoretical works on the
subject looks very short containing primarily \textit{ab initio}
approaches. Some experimentally established properties are fairly
well explained but some predictions are not confirmed by
experiment.

Already in the pioneer work \cite{tdio:Stashans96} a higher
possibility of lithium intercalation in the anatase structure than
in rutile was predicted as well as the absorption energies
obtained were calculated and Li-induced local one-electron energy
levels were found in the gap between the upper valence band and
the conduction band and could be attributed to Ti$^{3+}$ states.

The calculations of the relative lithium insertion potentials were
performed for the rutile, anatase, brookite, ramsdellite,
colombite, spinel, and orthorhombic polymorphs of titanium dioxide
from the first principles periodic Hartree-Fock approach
\cite{tdio:Mackrodt99} also indicating that lithium was completely
ionized in LiTiO$_2$ and that the charge transfer is predominantly
to the oxygen sublattice. A similar study of the average voltage
to intercalate lithium in various metal oxides (among them
TiO$_2$) and dichalcogenides was performed utilizing the
\textit{ab initio} pseudopotential method \cite{tdio:Aydinol97}.
It was also found that Li was fully ionized in the intercalated
compounds but with its charge distributed among the anion and the
metal.

A series of works
\cite{tdio:Koudriachova01,%
tdio:Koudriachova02,%
tdio:Koudriachova02b,%
tdio:Koudriachova02c,%
tdio:Koudriachova03,%
tdio:Koudriachova03b,%
tdio:Koudriachova04}
should be mentioned, where the lithium intercalation in both
rutile and anatase was \textit{ab initio} modelled taking into
account thermodynamic and kinetic effects.
The important role of strong local deformations of the lattice and
elastic screening of interlithium interactions was established,
the absence of insertion into rutile at room temperature was
explained in terms of inaccessibility of the low-energy
configurations due to highly anisotropic diffusion, a phase
separation in anatase into a Li-rich phase and a Li-poor phase was
described and the existence of a new distorted rock-salt phase for
LiTiO$_{2}$ was predicted
\cite{tdio:Koudriachova01,tdio:Koudriachova04}.
The calculated open circuit voltage profile reproduced and
explained the characteristic features of experimental discharge
curves for both polymorphs \cite{tdio:Koudriachova02b}.
An analysis of the site preference for Li intercalation in rutile
and diffusion pathways of ions was performed. The expansion of the
host structure on Li insertion was found to contribute to the
enhanced diffusion of Li ions along the $c$ direction while a
large distortion of the rutile framework nearly suppressed Li
diffusion in the $ab$ planes; computed diffusion coefficients were
found in excellent agreement with the measured values
\cite{tdio:Koudriachova03}.
A new phase of LiTiO$_{2}$ is predicted which may be accessed
through electrochemical lithiation of ramsdellite-structured
TiO$_{2}$ at the lowest potential (remaining constant over a wide
range of Li concentrations) reported for titanium dioxide based
materials \cite{tdio:Koudriachova08}.

\textit{Ab initio} study of the elastic properties of single and
polycrystal TiO$_{2}$ and other IV-B group oxides in the cotunnite
structure was performed \cite{tdio:Caravaca09}.

Up till now only one non-\textit{``ab initio''} description of the
intercalation in titanium dioxide by means of the
pseudospin-electron model \cite{tdio:Mysakovych07} (where the
pseudospin formalism was used in describing the intercalant
subsystem) was performed. The possibility of the phase transitions
accompanied by an abrupt change of the concentration of
intercalated ions and a significant increase of electrostatic
capacity of the system was predicted.

\subsection{Special features of the lithium intercalated anatase:
phase equilibrium and double positions for lithium ions}

Upon lithium insertion, an increasing fraction of the material
changes its crystallographic structure from anatase TiO$_{2}$ to
Li-rich lithium titanate Li$_{0.6}$TiO$_{2}$ (sometimes a
different stoichiometry is reported: Li$_{0.5}$TiO$_2$ or
Li$_{0.55}$TiO$_2$; as will be shown below it depends on the size
of TiO$_2$ crystallites). Phase separation occurs on the Li-rich
and the Li-poor (Li$_{0.01}$TiO$_2$) phases
\cite{tdio:Wagemaker01}.

Such a two-phase equilibrium system in the electrodes provides a
constant electrical potential between their electrodes (so-called
plateau in potential on the discharge curve) for a wide range of
the lithium concentration, because only the relative phase
fractions vary on charging (or discharging) the lithium while
their stoichiometries remain unchanged \cite{tdio:Wagemaker02}.
The Li-rich lithium titanate phase progressively moves inside the
anatase electrode as a front parallel to the interface and returns
during lithium extraction exactly in the way it came in
\cite{tdio:Wagemaker03b}.

Both in the anatase and in the lithium titanate lattice, Li is
found to be hopping over the available sites with activation
energies of 0.2 and 0.09~eV, respectively. However, macroscopic
intercalation data show activation energies of 0.5 eV because the
diffusion through the phase boundary determines the activation
energy of the overall diffusion and the overall diffusion rate
itself \cite{tdio:Wagemaker01}.

Recent NMR spectroscopy study \cite{tdio:Wagemaker07b} of
nanosized lithiated anatase revealed further important details of
the phase behavior and morphology. The coexistence of the Li-poor
and the Li-rich phases is possible only in the particles of the
size exceeding 120~nm due to the surface strain (occurring between
the phases) which becomes energetically unfavourable in small
particles. For the system of 40~nm particles, phase
stoichiometries are not stable indicating an enhanced solid
solution behavior: lithium content increases to Li$_{0.1}$TiO$_2$
in Li-poor and to Li$_{0.7}$TiO$_2$ in Li-rich phases. Further
decrease of the particle size makes it possible to find a fully
occupied phase Li$_{1}$TiO$_2$  \cite{tdio:Wagemaker07}. It can
coexist with the Li-rich one penetrating to the 3--4~nm depth
below the particle surface (or transforming all the particles less
than 7~nm in size). The poor Li ion conductivity can be due to the
full occupation of the octahedral voids, whereas ion diffusion
requires vacancies. Most likely, the short diffusion path in
nanostructured materials diminishes this problem as well as
elevation of temperature.

As established by quasi-elastic neutron scattering
\cite{tdio:Wagemaker03} Li ions can occupy two distinct positions
within the octahedral interstices along the $c$ axis (but only one
of them at a time). In the Li-anatase those positions are
symmetrical, separated by 1.61~\AA{} and equally occupied while in
the Li-titanate they are shifted, separated by only 0.7~\AA{} and
nonequivalent ($n_{\mathrm{Li1}}=0.32$ and $n_{\mathrm{Li2}}=0.19$
at 10~K -- the fitting of the site occupancy assuming a Boltzmann
distribution indicates that the energy difference between the
positions is 3.8~meV; positions 1 and 2 have an antiparallel
orientation in the neighbouring octahedra due to the phase
symmetry). A combination of quasi-elastic neutron scattering and
force field molecular dynamics simulations shows that Li is
hopping on a picosecond time scale between the two sites in the
octahedral interstices \cite{tdio:Wagemaker04b,tdio:Gubbens06}.

Lithium was also found to occupy multiple positions inside the
distorted oxygen octahedron of
Li$_{x}$Mg$_{0.1}$Ni$_{0.4}$Mn$_{1.5}$O$_{4}$ spinel
\cite{tdio:Wagemaker04}. Quite possible that this feature is
common to a wide family of crystals with a similar structure but
has been found only recently due to a higher precision of the
experiment.

\section{Symmetry analysis of the lithium intercalated anatase:
a possibility of internal piezoeffect}%
\label{sectn:SymmAnls}%

\begin{wrapfigure}[15]{i}{0.5\textwidth}
\vspace*{-4ex}
\centerline{\includegraphics[width=0.175\textwidth]{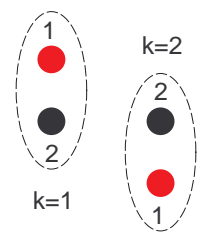}}
\vspace*{-1ex}
\caption{A schematic illustration of the positions available for
the intercalated lithium ion in oxygen octahedron voids of the
anatase elementary cell. Occupation of positions 1 and 2 is equal
in the poor Li-anatase phase and different in the rich Li-titanate
phase; their orientations are antiparallel in the neighbouring
octahedra (sublattices $k=1$, 2).}
\label{fig:positions}
\end{wrapfigure}
As mentioned before, the standard crystallographic cell of
body-centered anatase is chosen to consist of two elementary cells
for convenience. Thus, the respective space group $D_{4h}^{19}$
(or $I4_1/amd$ origin choice~2, No.~141) has a double set of
symmetry operations as compared with its point symmetry group
$D_{4h}$ ($4/mmm$). However, a single elementary cell obeys all
transformation rules. It consists of two formula units, i.e. two
octahedral voids (formed by oxygen anions) where the intercalated
lithium ion can reside in one of two available positions
(figure~\ref{fig:positions}).

Occupation of each octahedron by the lithium ion could be easily
described by the Hubbard projection operator $X_{ik}^{pp}$, where
$i$ is the lattice site index (i.e.\ the elementary cell index),
$k=1$, 2 is the sublattice index (the octahedron index), and the
state $p=0$ corresponds to an empty octahedron while the states
$p=1$, 2 denote the lithium ion in positions~1 or 2, respectively.
The Hubbard operator formalism reflects the microscopic structure
of the system and is very convenient for further calculations.

Alternatively, in the pseudospin formalism localization of lithium
in a certain position can be described by the pseudospin operator
$\hat{s}_{ik}=(-1)^{k-1}(X_{ik}^{11}-X_{ik}^{22})$. However, one
should also take into account the total occupation of the void
$\hat{n}_{ik}=X_{ik}^{11}+X_{ik}^{22}$. Such an approach separates
dipole-dipole (pseudospin) and particle-particle interactions in
the spirit of the Blume-Emery-Griffiths (BEG) model \cite{Blu71}.

Finally, symmetrized linear combination of the averages
\begin{equation}
    n_{\pm} = \frac{1}{2} (n_1 \pm n_2 ),
    \quad
    \eta_{\pm} = \frac{1}{2}(s_1 \pm s_2)
    \qquad
    \left(
        {n}_{k} \equiv \langle\hat{n}_{ik}\rangle,\;
        {s}_{k} \equiv \langle\hat{s}_{ik}\rangle
    \right)
\label{eq:symcmb}
\end{equation}
inherit symmetry properties of the system, thus being the order
parameters of possible phase transitions (see
appendix). Namely, $n_{+}$ (it transforms
according to the irreducible representation (IR)  ${A}_{1g}$ of
the point group $D_{4h}$) corresponds to the lithium concentration
(the average occupation of octahedral voids), $n_{-}$ (IR
${B}_{2u}$) is the difference of the void occupations in the
sublattices~1 and 2, $\eta_{+}$ (IR ${A}_{2u}$) is the
polarization along the $z$~axis, and $\eta_{-}$ (IR ${B}_{1g}$
which corresponds to the phase transition into the Li-titanate
point symmetry subgroup $D_{2h}$) simultaneously describes two
phenomena: the antipolarization along the $z$~axis (unlike the
true antiferroelectric ordering with doubling of the unit cell, a
mutual compensation of sublattice polarizations occurs here just
as in the Mitsui model) and the deformation $U_{xx}-U_{yy}$ in the
$ab$~plane. Such a coexistence of the antipolar ordering and the
deformation belonging to the same irreducible representation and,
thus, described by a common order parameter is called internal
piezoeffect.

\section{Phase equilibrium in the framework
of the Landau expansion}

The symmetry analysis performed in the previous section can serve
as a background for a qualitative description of thermodynamics of
the considered system in the framework of Landau expansion
\begin{equation}
    F = F_0 + \frac{1}{2} a \rho^2 + \frac{1}{3} b \rho^3
        + \frac{1}{4} c \rho^4
        + \frac{1}{2} A \eta^2 + \frac{1}{4} B \eta^4
        - \sigma \eta - \mu \rho \,,
\label{eq:LandauExp}
\end{equation}
where $\rho$ describes the intercalant (lithium) concentration and
hence corresponds to the $n_{+}$ introduced above, $\mu$ is the
chemical potential, the order parameter $\eta$ is proportional to
the $\eta_{-}$, and the conjugated ``field'' $\sigma$ describes
the applied stress; the expansion coefficients should satisfy the
following conditions
\[
A = A_0 + A_1 \rho, \qquad B > 0, \qquad c > 0.
\]
The equilibrium state of the system is achieved at the minimum of
the free energy
\begin{align}
    \frac{\partial F}{\partial \rho} &= a \rho + b \rho^2 + c \rho^3
        + \frac{1}{2} A_1 \eta^2 - \mu = 0,
\label{eq:RhoEqlbr}
\\%
    \frac{\partial F}{\partial \eta} &= A \eta + B \eta^3 - \sigma = 0.
\label{eq:EtaEqlbr}
\end{align}

Further calculations are limited to the case of zero ``field'':
$\sigma=0$. Then, equation \eqref{eq:EtaEqlbr} can have either
trivial solution $\eta=0$ or nonzero one:
\begin{equation}
    A + B \eta_0^2 = 0 \quad \Rightarrow \quad
        \eta_0 = \pm \sqrt{-(A_0 + A_1\rho)/B}.
\label{eq:Eta0}
\end{equation}
At $B>0$, $\eta_0$ takes on a real value under the condition $A_0
+ A_1\rho < 0$ which means $A_0>0$ and $A_1<0$ giving
\begin{equation}
    \eta_0^2 = \frac{\abs{A_1}\rho - A_0}{B},\qquad \rho > \frac{A_0}{\abs{A_1}}.
\label{eq:Eta02}
\end{equation}
Let us consider the cases $\eta=0$ and $\eta\neq0$ separately.

\begin{itemize}
\item[I.] $\eta=0$\\%
As follows from equation \eqref{eq:RhoEqlbr}
\begin{equation}
    \varphi(\rho) = \mu,\qquad
        \varphi(\rho) \equiv a \rho + b \rho^2 + c \rho^3.
\label{eq:EtaZeroEq}
\end{equation}
This equation could have three solutions in a certain region of
chemical potential values (i.e.\ the possibility of a phase
transition with the jump of $\rho$) if the extrema of the function
$\varphi(\rho)$ exist, i.e.\ the equation
\[
    \frac{\partial \varphi(\rho)}{\partial \rho} =
        a + 2 b \rho + 3 c \rho^2 = 0
\]
has nonzero solutions
\begin{equation}
    \rho_{1,2}=\frac{1}{3c}\left[-b \pm \sqrt{b^2 - 3 a c}\, \right],
\label{eq:RhoSoltns}
\end{equation}
which imposes a condition on the Landau expansion coefficients
\begin{equation}
    b^2 - 3 a c > 0.
\label{eq:EtaZeroCnd}
\end{equation}
An equivalent condition could be obtained by setting the second
derivative to zero
\[
    \frac{\partial^2 \varphi(\rho)}{\partial \rho^2} =
        2 b + 6 c \rho = 0,
\]
which gives the ordinate of the inflection point
\begin{equation}
    \rho^{\ast} = - b / 3 c,
\label{eq:EtaZeroCPnt}
\end{equation}
and demanding a negative value of the first derivative at this
point
${\partial\varphi(\rho)}/{\partial\rho}|_{\rho=\rho^{\ast}}<0$.
Since the curve $\varphi(\rho)$ always crosses the inflection
point (which is the symmetry centre of the curve), this point is
also crossed by the line of the phase transition occurring at the
following value of chemical potential
\begin{equation}
    \varphi(\rho^{\ast}) = \frac{b}{3c}\left(\frac{2b^2}{9c}-a\right)
        = \mu^{\ast}.
\label{eq:EtaZeroCMu}
\end{equation}
Considering that the parameter $\rho$ describes concentration, we
have an additional condition $\rho\geqslant 0$: both solutions of
equation \eqref{eq:RhoSoltns} are positive if $b>0$.
\item[II.] $\eta=\pm\eta_0$\\%
After the identical calculations one can obtain an expression
similar to equation \eqref{eq:EtaZeroEq} but slightly
renormalized:
\begin{equation}
    \tilde{a} \rho + b \rho^2 + c \rho^3 = \tilde{\mu},\qquad
        \tilde{a} = a - \frac{\abs{A_1}^2}{2B},\quad
        \tilde{\mu} = \mu - \frac{A_0\abs{A_1}}{2B}.
\label{eq:EtaNZeroEq}
\end{equation}
The phase transition exists if
\begin{equation}
    b^2 - 3 \tilde{a} c > 0
\label{eq:EtaNZeroCnd}
\end{equation}
and it occurs at the following value of chemical potential
\begin{equation}
    \tilde{\mu}^{\ast} = \frac{b}{3c}\left(\frac{2b^2}{9c}-\tilde{a}\right).
\label{eq:EtaNZeroCMu}
\end{equation}
\end{itemize}

As follows from the above considerations, the behaviour of the
system at change of the chemical potential depends on the values
of Landau expansion coefficients. So, further considerations are
limited to the case which qualitatively describes the phase
transition between the poor phase I ($\eta=0$, $\rho\to0$) and the
rich phase II ($\eta\neq0$, $\rho\to0.5$) in the lithiated
anatase.

Comparing conditions \eqref{eq:EtaZeroCnd} and
\eqref{eq:EtaNZeroCnd} one can derive the condition
\begin{equation}
    3 a c - \frac{3}{2} \frac{\abs{A_1}^2}{B}c < b^2 < 3 a c
\label{eq:EtaCombCnd}
\end{equation}
describing the case when only one solution for $\rho$ exists in
the phase~I while there are three possible solutions in the
phase~II at nonzero $\eta_0$. Combining the equations for
equilibrium values of $\rho$ and the condition of the first order
phase transition
$F_{\text{I}}=F_{\text{II}}$
between the phases~I and~II, we obtain a set of equations for
values of $\rho_{\text{I}}$, $\rho_{\text{II}}$ and $\mu$ at the
phase transition point
\begin{align}
    & a \rho_{\text{I}} + b \rho_{\text{I}}^2 + c \rho_{\text{I}}^3 = \mu,
\notag\\%
    & \tilde{a} \rho_{\text{II}} + b \rho_{\text{II}}^2 + c \rho_{\text{II}}^3 = \tilde{\mu},
\notag\\%
    & \frac{1}{2} a \rho_{\text{I}}^2 + \frac{1}{3} b \rho_{\text{I}}^3
        + \frac{1}{4} c \rho_{\text{I}}^4
        - \mu \rho_{\text{I}} =
\notag\\%
    & \quad = \frac{1}{2} a \rho_{\text{II}}^2 + \frac{1}{3} b \rho_{\text{II}}^3
        + \frac{1}{4} c \rho_{\text{II}}^4
        + \frac{1}{2} A(\rho_{\text{II}}) \eta_0^2(\rho_{\text{II}})
        + \frac{1}{4} B \eta_0^4(\rho_{\text{II}})
        - \mu \rho_{\text{II}}.
\label{eq:PhTrPnt}
\end{align}

\begin{wrapfigure}{i}{0.5\textwidth}
\centerline{\includegraphics[width=0.47\textwidth]{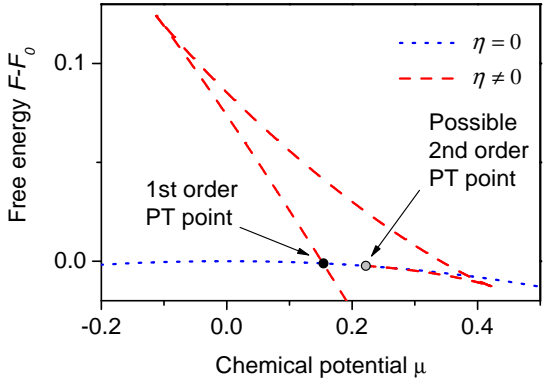}}
\caption{Dependence of free energy on chemical potential for
branches with zero and nonzero order parameter $\eta$: the first
order phase transition takes place.}
\label{fig:FrEng}
\end{wrapfigure}
A qualitative illustration of such a phase transition, which
corresponds to the case in the lithium intercalated anatase, is
given in figure~\ref{fig:FrEng} for the following set of Landau
coefficients \eqref{eq:LandauExp}: $a=10.8$, $b=-35.2$, $c=45.7$,
$A_0=0.2$, $A_1=-9.1$, $B=9$. This set satisfies the condition
\eqref{eq:EtaCombCnd} and, as follows from the free energy
analysis, at the rise of chemical potential the first order phase
transition from the branch~I to the branch~II \eqref{eq:PhTrPnt}
precedes the possible second order phase transition with a
continuous growth of the nonzero value of $\rho$ at
$\rho(\mu)=\rho_c$.

\begin{figure}
\includegraphics[width=0.47\textwidth]{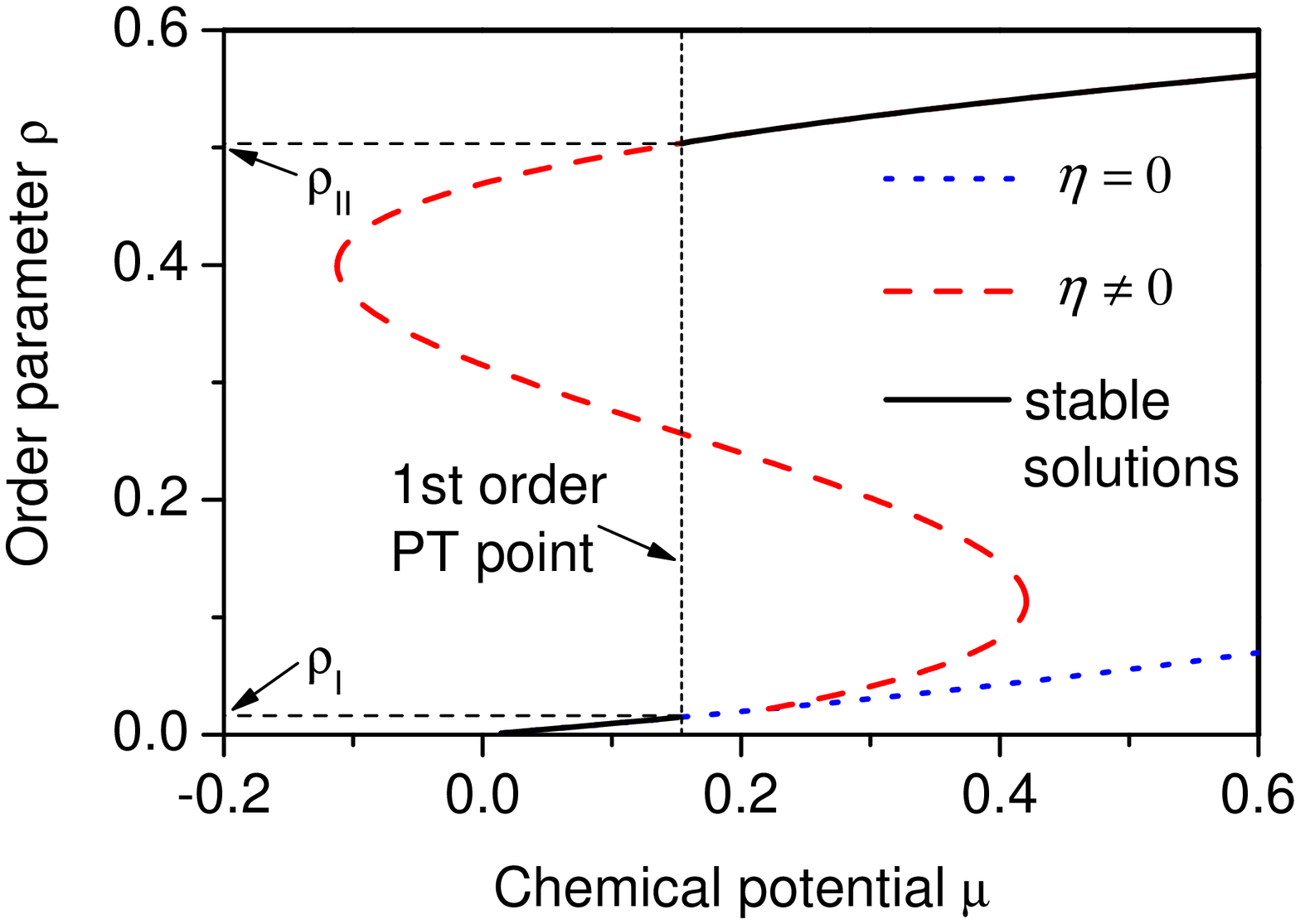}%
\hfill%
\includegraphics[width=0.47\textwidth]{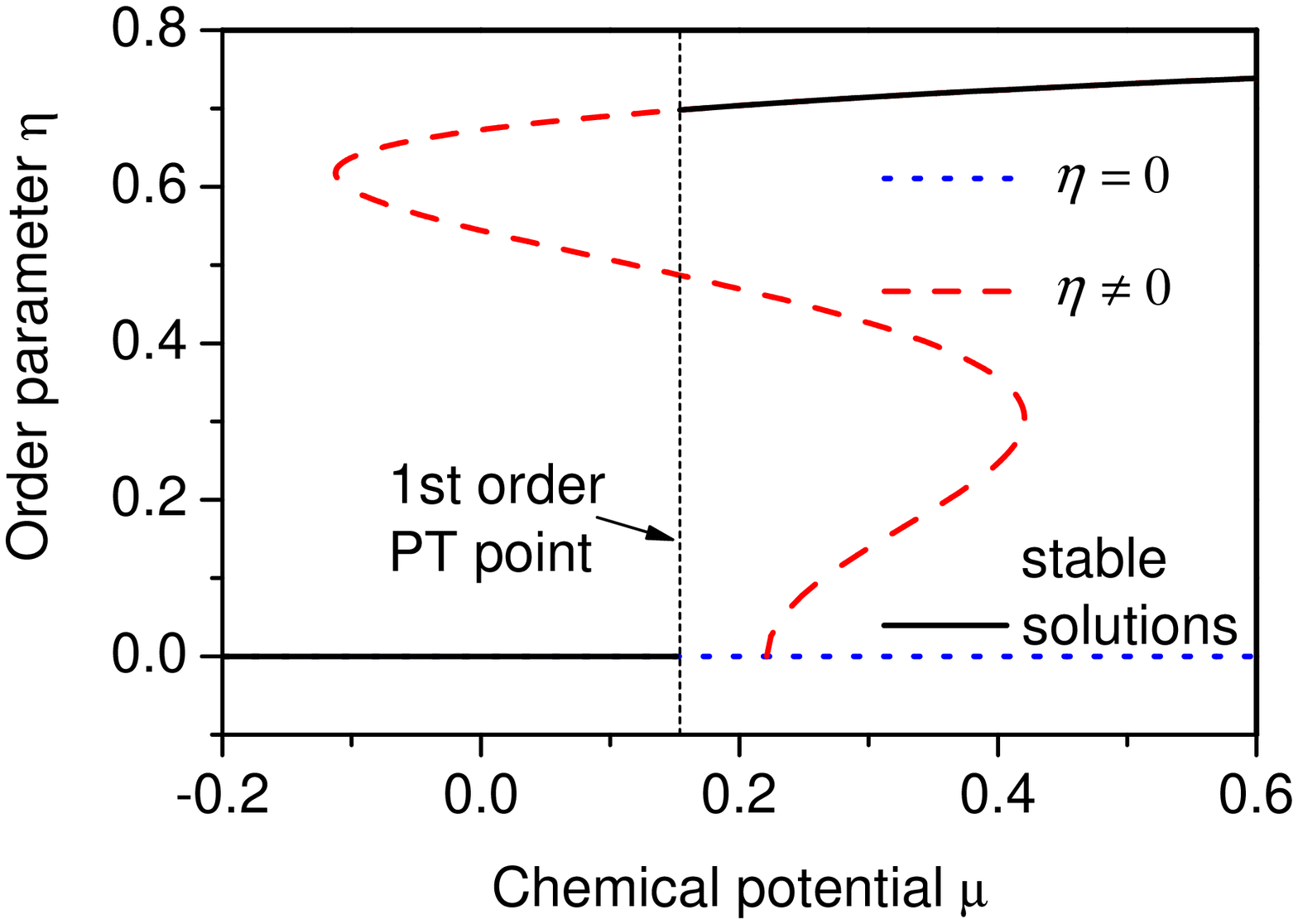}%
\caption{A jump of order parameters $\rho$ (left) and $\eta$
(right) at the first order phase transition. Thermodynamically
stable solutions are marked with the solid curve. In the
$\rho=\text{const}$ regime the phase separation into phases with
$\rho_{\text{I}}$ and $\rho_{\text{II}}$ takes place.}
\label{fig:RhoEta}
\end{figure}

Due to the dependence on parameter $\rho$ of the coefficients of
parameter $\eta$ both of them have simultaneous jumps at the point
of the first order phase transition (figure~\ref{fig:RhoEta}).
Parameter $\eta$ should be considered as a true order parameter of
this transition because it is exactly equal to zero in the initial
phase.
It should be also noted that (due to proximity of the phase
transition points) at increase of chemical potential the system
could ``pass through'' the first order phase transition point (a
metastable state) and the second order phase transition does occur
followed by the first order phase transition in the extremum point
of $\rho(\mu)$. All the above considerations are valid for the
case $\mu=\text{const}$. However, the lithiated anatase
corresponds rather to the system with the fixed lithium
concentration ($\rho=\text{const}$). In this case the system
separates into phases with concentrations $\rho_{\text{I}}$ and
$\rho_{\text{II}}$ (figure~\ref{fig:RhoEta}) and respective
weights $w_{\text{I}}$ and $w_{\text{II}}$, so
$\rho_{\text{fixed}}=w_{\text{I}}\rho_{\text{I}}+w_{\text{II}}\rho_{\text{II}}$.
The chemical potential of the system is constant and is equal to
the chemical potential value at which the first order phase
transition occurs in the $\mu=\text{const}$ regime.

Using the equality
\[
    \frac{\partial^2 F}{\partial \eta^2}= A + 3 B \eta^2
        = \frac{\partial \sigma}{\partial \eta},
\]
one can calculate the susceptibility which describes the reaction
of the order parameter $\eta$ (the deformation $U_{xx}-U_{yy}$)
with respect to the ``field'' $\sigma$ (this susceptibility is
related to the elastic modulus of the system)
\begin{equation}
    \chi \equiv \frac{\partial \eta}{\partial \sigma}
        = \left[A_0 - \abs{A_1} \rho + 3 B \eta^2 \right]^{-1}.
\label{eq:LSuscpt}
\end{equation}
Its explicit form depends on the phase
\begin{align}
    \eta = 0 &\colon\quad
        \chi_{\text{I}} =
        \left[A_0-\abs{A_1}\rho_{\text{I}}\right]^{-1},
\label{eq:LSuscptI}
\\%
    \eta = \eta_0 &\colon\quad
        \chi_{\text{II}} =
        \frac{1}{2}\left[\abs{A_1}\rho_{\text{II}}-A_0\right]^{-1}.
\label{eq:LSuscptJump}
\end{align}
Due to the phase transition, the susceptibility has a jump whose
value can be calculated from expressions \eqref{eq:LSuscptI} and
\eqref{eq:LSuscptJump} using solutions of the set
\eqref{eq:PhTrPnt}.

Utilizing the equality $A_0 - \abs{A_1} \rho = -B \eta^2$, one can
rewrite the susceptibility \eqref{eq:LSuscpt} at $\eta\neq0$ in
the form
\begin{equation}
    \chi = \left[2 B \eta^2 \right]^{-1}.
\label{eq:LSuscptEta}
\end{equation}
Thus, the susceptibility could diverge in both cases
\eqref{eq:LSuscptI} and \eqref{eq:LSuscptEta} approaching the
point of the possible second order phase transition from the
respective direction.

\section{Lattice model for the lithium intercalated anatase}

\subsection{Model Hamiltonian and thermodynamics
in the mean field approximation}

As demonstrated in the previous section, the Landau expansion
combined with the symmetry analysis gives a good qualitative
picture of the phase separation in the lithiated anatase. However,
a detailed description of temperature dependent thermodynamic
properties of the system could be derived only in the framework of
a microscopic approach. Let us construct a model Hamiltonian of
the lattice gas type:
\begin{equation}
    \hat{H} = \hat{H}_1+\hat{H}_{\text{int}}+\hat{H}_{\text{def}},
\label{eq:ModHamltn}
\end{equation}
where besides the single particle term $\hat{H}_1$ the
interparticle interaction $\hat{H}_{\text{int}}$ and the lattice
deformation $\hat{H}_{\text{def}}$ are taken into account:
\begin{align}
    \hat{H}_1 &= \sum_i \sum_k \sum_p (\varepsilon_{0}-\mu) X_{ik}^{pp}
    - h \sum_i (s_{i1}{+}s_{i2})
    - \Delta \sum_i (s_{i1}{-}s_{i2}),
\notag\\%
    \hat{H}_{\text{int}} &= -\frac{1}{2} \sum_{i\neq j} \sum_{kl}
    \sum_{pq} W_{kl}^{pq}(i,j) X_{ik}^{pp} X_{jl}^{qq},
\notag\\%
    \hat{H}_{\text{def}} &= \frac{1}{2} NCU^2,
\end{align}
where
\[
    \Delta=\alpha U,\quad U = U_{xx}-U_{yy};\qquad
    k,l=1,2,\qquad p,q=0,1,2;
\]
$\mu$ is chemical potential of the intercalant particles (let us
assign $\varepsilon_{0}$ as its origin), $h$ is the external
electric field, $\Delta$ is the deformation induced internal
field, $U$ is the effective deformation in the $XY$ plane
($\sigma=\partial(\Omega/N)/\partial U$ defines the stress),
$W_{kl}^{pq}(i,j)$ are the interaction energies between the
particles in the respective positions. Hence, we take into account
both the semiphenomenological deformational shift of the lattice
energy $\hat{H}_{\text{def}}$ and the effective internal staggered
field $\Delta$ (similar to the one in the Mitsui model) which
appears due to the intercalation induced lattice deformation
making the lithium intercalation positions inequivalent in the
pairs. Unlike the ordinary lattice gas approach, the proposed
model considers two equilibrium positions for intercalated
particles. Such multistate models (see, e.g.\
\cite{tdio:McKinnon83,tdio:Grygorchak07}) are rather rare since
monopositional intercalated materials are the most common.
Deformational effects have been also taken into account because an
effective potential for the Li ions was affected by the
intercalation-induced distortion of the host
\cite{tdio:Vakarin00}.

In the mean field approximation, Hamiltonian \eqref{eq:ModHamltn}
becomes linear
\begin{equation}
    \hat{H}_{\text{MFA}} = N E_0+ \sum_i \sum_k \sum_p H_{kp}
    X_{ik}^{pp}.
\label{eq:MFAHamltn}%
\end{equation}
Taking into account the symmetry properties of the Fourier
transforms of the interaction energies in the centre of the
Brillouin zone
\begin{align}
    W_{11}^{11} &= W_{11}^{22} = W_{22}^{11} = W_{22}^{22},&
    W_{11}^{12}  = W_{11}^{21} = W_{22}^{12} = W_{22}^{21},
\notag\\%
    W_{12}^{12} &= W_{21}^{21} = W_{12}^{21} = W_{21}^{12},&
    W_{12}^{11}  = W_{12}^{22} = W_{21}^{11} = W_{21}^{22},
\notag%
\end{align}
one can write down the expressions for average values and
effective fields
\begin{align}
    E_0 &= {\textstyle\frac{1}{2}} W_{11}^{11}
    \left(
    \langle X_1^{11}\rangle^2+\langle X_1^{22}\rangle^2
    +\langle X_2^{11}\rangle^2+\langle X_2^{22}\rangle^2
    \right)
\notag\\%
    &\quad{}
    +W_{11}^{12}
    \left(
    \langle X_1^{11}\rangle \langle X_1^{22}\rangle
    +\langle X_2^{11}\rangle \langle X_2^{22}\rangle
    \right)
\notag\\%
    &\quad{}
    +W_{12}^{11}
    \left(
    \langle X_1^{11}\rangle \langle X_2^{11}\rangle
    +\langle X_1^{22}\rangle \langle X_2^{22}\rangle
    \right)
\notag\\%
    &\quad{}
    +W_{12}^{12}
    \left(
    \langle X_1^{11}\rangle \langle X_2^{22}\rangle
    +\langle X_1^{22}\rangle \langle X_2^{11}\rangle
    \right),
\notag\\%
    H_{11} &= -\mu-h-\Delta
\notag\\%
    &\quad{}
    -\left(
    W_{11}^{11} \langle X_1^{11}\rangle + W_{11}^{12} \langle X_1^{22}\rangle
    +W_{12}^{11} \langle X_2^{11}\rangle + W_{12}^{12} \langle X_2^{22}\rangle
    \right),
\notag\\%
    H_{12} &= -\mu+h+\Delta
\notag\\%
    &\quad{}
    -\left(
    W_{11}^{11} \langle X_1^{22}\rangle + W_{11}^{12} \langle X_1^{11}\rangle
    +W_{12}^{11} \langle X_2^{22}\rangle + W_{12}^{12} \langle X_2^{11}\rangle
    \right),
\notag\\%
    H_{21} &= -\mu+h-\Delta
\notag\\%
    &\quad{}
    -\left(
    W_{11}^{11} \langle X_2^{11}\rangle + W_{11}^{12} \langle X_2^{22}\rangle
    +W_{12}^{11} \langle X_1^{11}\rangle + W_{12}^{12} \langle X_1^{22}\rangle
    \right),
\notag\\%
    H_{22} &= -\mu-h+\Delta
\notag\\%
    &\quad{}
    -\left(
    W_{11}^{11} \langle X_2^{22}\rangle + W_{11}^{12} \langle X_2^{11}\rangle
    +W_{12}^{11} \langle X_1^{22}\rangle + W_{12}^{12} \langle X_1^{11}\rangle
    \right),
\label{eq:EffFldX}%
\end{align}

Average occupations of the positions can be obtained as solutions
of the selfconsistency equation set
\begin{equation}
    \langle X_k^{pp}\rangle = Z_k^{-1} \mathrm{e}^{-\beta H_{kp}},
\label{eq:SelfcEqX}%
\end{equation}
where partition functions of sublattices are as follows:
\begin{equation}
    Z_k = 1 + \mathrm{e}^{-\beta H_{k1}} + \mathrm{e}^{-\beta H_{k2}}
\label{eq:Zk}%
\end{equation}
and thermodynamically stable solutions are chosen according to the
criterion of the minimum of grand canonical potential
\begin{equation}
    \Omega/N = E_0 + {\textstyle\frac{1}{2}} C U^2 - \Theta \ln (Z_1 Z_2).
\label{eq:GP1}%
\end{equation}

For the $(n,s)$-representation (introduced in
section~\ref{sectn:SymmAnls}) selfconsistency equations look like
\begin{align}
    n_k &= Z_k^{-1}
    \left(
    \mathrm{e}^{-\beta H_{k1}} + \mathrm{e}^{-\beta H_{k2}}
    \right),
\notag\\%
    s_k &= (-1)^{k-1} Z_k^{-1}
    \left(
    \mathrm{e}^{-\beta H_{k1}} - \mathrm{e}^{-\beta H_{k2}}
    \right),
\label{eq:SelfcNS}%
\end{align}
where the term $E_0$ and the effective fields are as follows
\begin{align}
    E_0 &= {\textstyle\frac{1}{4}}
    \bigl[
    W_{11}^{+}(n_1^2+n_2^2)+W_{11}^{-}(s_1^2+s_2^2)
    +2W_{12}^{+} n_1 n_2 - 2W_{12}^{-} s_1 s_2
    \bigr],
\notag\\%
    H_{11} &= -\mu-h-\Delta
    -{\textstyle\frac{1}{2}}
    \left(
    W_{11}^{+} n_1 + W_{11}^{-} s_1 + W_{12}^{+} n_2 - W_{12}^{-} s_2
    \right),
\notag\\%
    H_{12} &= -\mu+h+\Delta
    -{\textstyle\frac{1}{2}}
    \left(
    W_{11}^{+} n_1 - W_{11}^{-} s_1 + W_{12}^{+} n_2 + W_{12}^{-} s_2
    \right),
\notag\\%
    H_{21} &= -\mu+h-\Delta
    -{\textstyle\frac{1}{2}}
    \left(
    W_{12}^{+} n_1 + W_{12}^{-} s_1 + W_{11}^{+} n_2 - W_{11}^{-} s_2
    \right),
\notag\\%
    H_{22} &= -\mu-h+\Delta
    -{\textstyle\frac{1}{2}}
    \left(
    W_{12}^{+} n_1 - W_{12}^{-} s_1 + W_{11}^{+} n_2 + W_{11}^{-} s_2
    \right),
\label{eq:EffFldNS}%
\end{align}
and new combinations of interaction energies are introduced
\begin{equation}
    W_{11}^{\pm} = W_{11}^{11} \pm W_{11}^{12},\qquad
    W_{12}^{\pm} = W_{12}^{11} \pm W_{12}^{12}.
\label{eq:WNS}%
\end{equation}

Finally, the symmetrized combinations of averages
\eqref{eq:symcmb} (which can be the order parameters of the
system) are found from the following set
\begin{align}
n_{\pm} &= \frac{1}{2}
    \left[
        Z_1^{-1}
        \left(
            \mathrm{e}^{-\beta H_{11}}+\mathrm{e}^{-\beta H_{12}}
        \right)
        \pm
        Z_2^{-1}
        \left(
            \mathrm{e}^{-\beta H_{21}}+\mathrm{e}^{-\beta H_{22}}
        \right)
    \right],
\notag\\%
\eta_{\pm} &= \frac{1}{2}
    \left[
        Z_1^{-1}
        \left(
            \mathrm{e}^{-\beta H_{11}}-\mathrm{e}^{-\beta H_{12}}
        \right)
        \mp
        Z_2^{-1}
        \left(
            \mathrm{e}^{-\beta H_{21}}-\mathrm{e}^{-\beta H_{22}}
        \right)
    \right],
\label{eq:SelfcOP}%
\end{align}
with the respective definitions for $E_0$ and the effective fields
\begin{align}
    E_0 &= {\textstyle\frac{1}{2}}
    \bigl[
    W_{++} n_{+}^2 + W_{+-} n_{-}^2
    + W_{--} \eta_{+}^2 + W_{-+} \eta_{-}^2
    \bigr],
\notag\\%
    H_{11} &= -\mu-h-\Delta
    -{\textstyle\frac{1}{2}}
    \left(
    W_{++} n_{+} + W_{+-} n_{-} + W_{--} \eta_{+} + W_{-+} \eta_{-}
    \right),
\notag\\%
    H_{12} &= -\mu+h+\Delta
    -{\textstyle\frac{1}{2}}
    \left(
    W_{++} n_{+} + W_{+-} n_{-} - W_{--} \eta_{+} - W_{-+} \eta_{-}
    \right),
\notag\\%
    H_{21} &= -\mu+h-\Delta
    -{\textstyle\frac{1}{2}}
    \left(
    W_{++} n_{+} - W_{+-} n_{-} - W_{--} \eta_{+} + W_{-+} \eta_{-}
    \right),
\notag\\%
    H_{22} &= -\mu-h+\Delta
    -{\textstyle\frac{1}{2}}
    \left(
    W_{++} n_{+} - W_{+-} n_{-} + W_{--} \eta_{+} - W_{-+} \eta_{-}
    \right),
\label{eq:EffFldOP}%
\end{align}
as well as symmetrized interaction energies
\begin{equation}
    W_{+\pm} = W_{11}^{+} \pm W_{12}^{+},\qquad
    W_{-\pm} = W_{11}^{-} \pm W_{12}^{-}.
\label{eq:WOP}%
\end{equation}

Considering the definition $\sigma=\partial(\Omega/N)/\partial U$
with account of the equality $\Delta=\alpha U$ and the expression
for grand canonical potential \eqref{eq:GP1}, the equation for the
deformation is obtained
\begin{equation}
    U = \frac{2\alpha}{C}(\tilde{\sigma}+\eta_{-}),
\label{eq:UfuncSE}%
\end{equation}
where $\tilde{\sigma}=\sigma/2\alpha$ is a scaled dimensionless
stress. As it follows
\begin{equation}
    \frac{1}{2} CU^2 = k_{\Delta}(\tilde{\sigma}+\eta_{-})^2,
    \qquad
    \Delta = k_{\Delta}(\tilde{\sigma}+\eta_{-}),
\label{eq:kDlt}%
\end{equation}
where $k_{\Delta}={2\alpha^2}/{C}$. Thus, the deformation
\eqref{eq:UfuncSE} can occur spontaneously (giving rise to the
antisymmetrical internal field $\Delta$) due to the appearance of
the order parameter $\eta_{-}$ even at the absence of the stress.

The deformation $U$ is a proper variable for the grand canonical
potential $\Omega$ \eqref{eq:GP1} but in our case it is convenient
to deal with the stress $\sigma$ (conjugated to $U$). Performing
the Legendre transformation
\[
    \mathrm{d}\Omega = \dotsb + \sigma \mathrm{d} U
            =  \dotsb + \mathrm{d} (\sigma U) - U \mathrm{d} \sigma,
\]
one can build the desired form of grand canonical potential
$\widetilde{\Omega}$
\begin{equation}
    \widetilde{\Omega} = \Omega - \sigma U.
\label{eq:tldOmg}%
\end{equation}
The respective deformational term of the thermodynamic potential
looks like
\[
    \frac{1}{2} CU^2 - \sigma U =
        k_{\Delta}(\eta_{-}^2-\tilde{\sigma}^2).
\]

\subsection{Phase diagram of the ground state}

At zero temperature, the homogenious system, which is described by
Hamiltonian \eqref{eq:ModHamltn}, could reside in one of the nine
possible states $\lvert p_1 p_2 \rangle$ (let us also use a more
descriptive notation where ``up'' and ``down'' arrows indicate
occupied positions, e.g.\ $\lvert 10 \rangle\equiv\lvert
{\uparrow} 0 \rangle$) with the following values of
thermodynamic potential \eqref{eq:tldOmg}:
\begin{align}
    \lvert 00 \rangle&\equiv\lvert 0 \rangle\colon&
    \widetilde{\Omega}_0 &=- k_{\Delta}\tilde{\sigma}^2,&&&&
\notag\\%
    \lvert {\uparrow} 0 \rangle&\equiv\lvert 1 \rangle\colon&
    \widetilde{\Omega}_1 &\textstyle= -\mu -h -W_1 - k_{\Delta}(\tilde{\sigma}+\frac{1}{2})^2,&&&&
\notag\\%
    \lvert {\downarrow} 0 \rangle&\equiv\lvert 2 \rangle\colon&
    \widetilde{\Omega}_2 &\textstyle= -\mu +h -W_1 - k_{\Delta}(\tilde{\sigma}-\frac{1}{2})^2,&&&&
\notag\\%
    \lvert 0 {\downarrow} \rangle&\equiv\lvert 3 \rangle\colon&
    \widetilde{\Omega}_3 &\textstyle= -\mu +h -W_1 - k_{\Delta}(\tilde{\sigma}+\frac{1}{2})^2,&&&&
\notag\\%
    \lvert 0 {\uparrow} \rangle&\equiv\lvert 4 \rangle\colon&
    \widetilde{\Omega}_4 &\textstyle= -\mu -h -W_1 - k_{\Delta}(\tilde{\sigma}-\frac{1}{2})^2,&&&&
\notag\\%
    \lvert {\uparrow} {\uparrow} \rangle&\equiv\lvert 5 \rangle\colon&
    \widetilde{\Omega}_5 &\textstyle= -2\mu -2h -2W_2 - k_{\Delta}\tilde{\sigma}^2,&&&&
\notag\\%
    \lvert {\downarrow} {\downarrow} \rangle&\equiv\lvert 6 \rangle\colon&
    \widetilde{\Omega}_6 &\textstyle= -2\mu +2h -2W_2 - k_{\Delta}\tilde{\sigma}^2,&&&&
\notag\\%
    \lvert {\uparrow} {\downarrow} \rangle&\equiv\lvert 7 \rangle\colon&
    \widetilde{\Omega}_7 &\textstyle= -2\mu -2W_3 - k_{\Delta}(\tilde{\sigma}+1)^2,&&&&
\notag\\%
    \lvert {\downarrow} {\uparrow} \rangle&\equiv\lvert 8 \rangle\colon&
    \widetilde{\Omega}_8 &\textstyle= -2\mu -2W_3 - k_{\Delta}(\tilde{\sigma}-1)^2;&&&&
\label{eq:GStldOmg}%
\end{align}
where
\begin{align}
    W_1 &\textstyle= \frac{1}{8} (W_{++}+W_{+-}+W_{--}+W_{-+})
        = \frac{1}{2} W_{11}^{11},
\notag\\%
    W_2 &\textstyle= \frac{1}{4} (W_{++}+W_{--})
        = \frac{1}{2} (W_{11}^{11}+W_{12}^{12}),
\notag\\%
    W_3 &\textstyle= \frac{1}{4} (W_{++}+W_{-+})
        = \frac{1}{2} (W_{11}^{11}+W_{12}^{11}).
\label{eq:W123}%
\end{align}
As is obvious from expressions \eqref{eq:GStldOmg}, at
$\tilde{\sigma}>0$ (in particular at
$\tilde{\sigma}\to+\varepsilon$) the levels 1, 3, and 7
\emph{always} lie below the levels 4, 2 and 8, respectively. So, the
latter will not be further considered.

\begin{figure}
\includegraphics[width=0.47\textwidth]{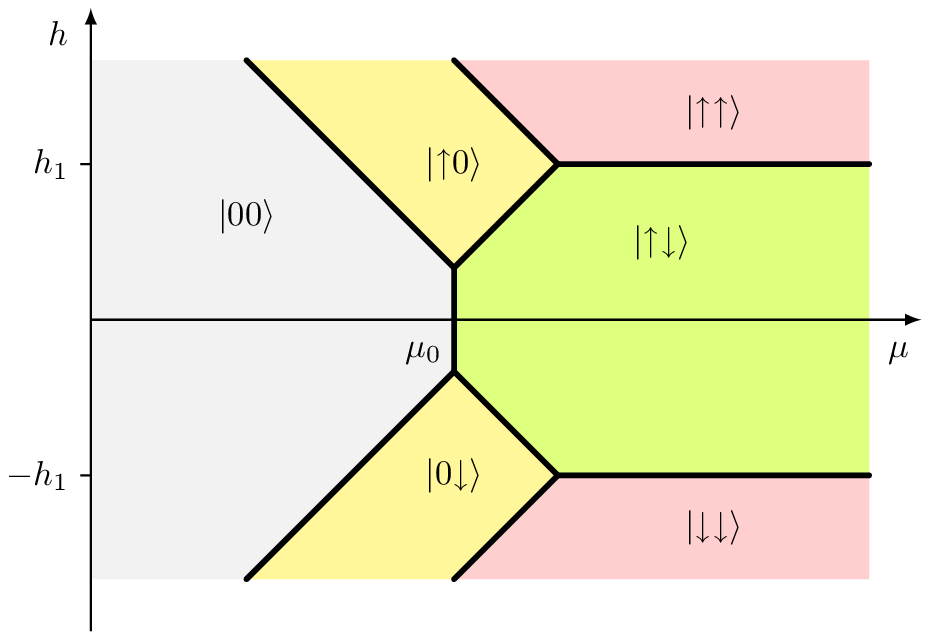}%
\hfill%
\includegraphics[width=0.47\textwidth]{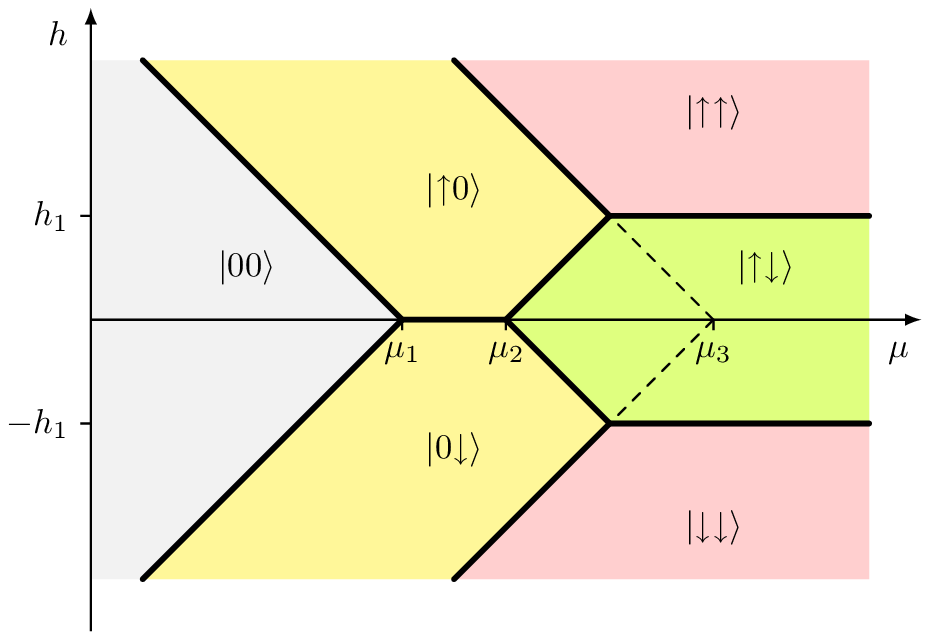}%
\\%
\parbox[t]{0.47\textwidth}{%
\caption{The most general form of the ground state phase diagram:
existence of intermediate half-filled phases and the possibility
of a direct transition from the empty state to the full one.}
\label{fig:GSPD-C}%
}%
\hfill%
\parbox[t]{0.47\textwidth}{%
\caption{The ground state phase diagram with the phase transition
between the empty and half-filled states which corresponds to the
Li-poor -- Li-rich phase equilibrium in the lithiated anatase.}
\label{fig:GSPD-A}%
}
\end{figure}

\begin{figure}
\includegraphics[width=0.47\textwidth]{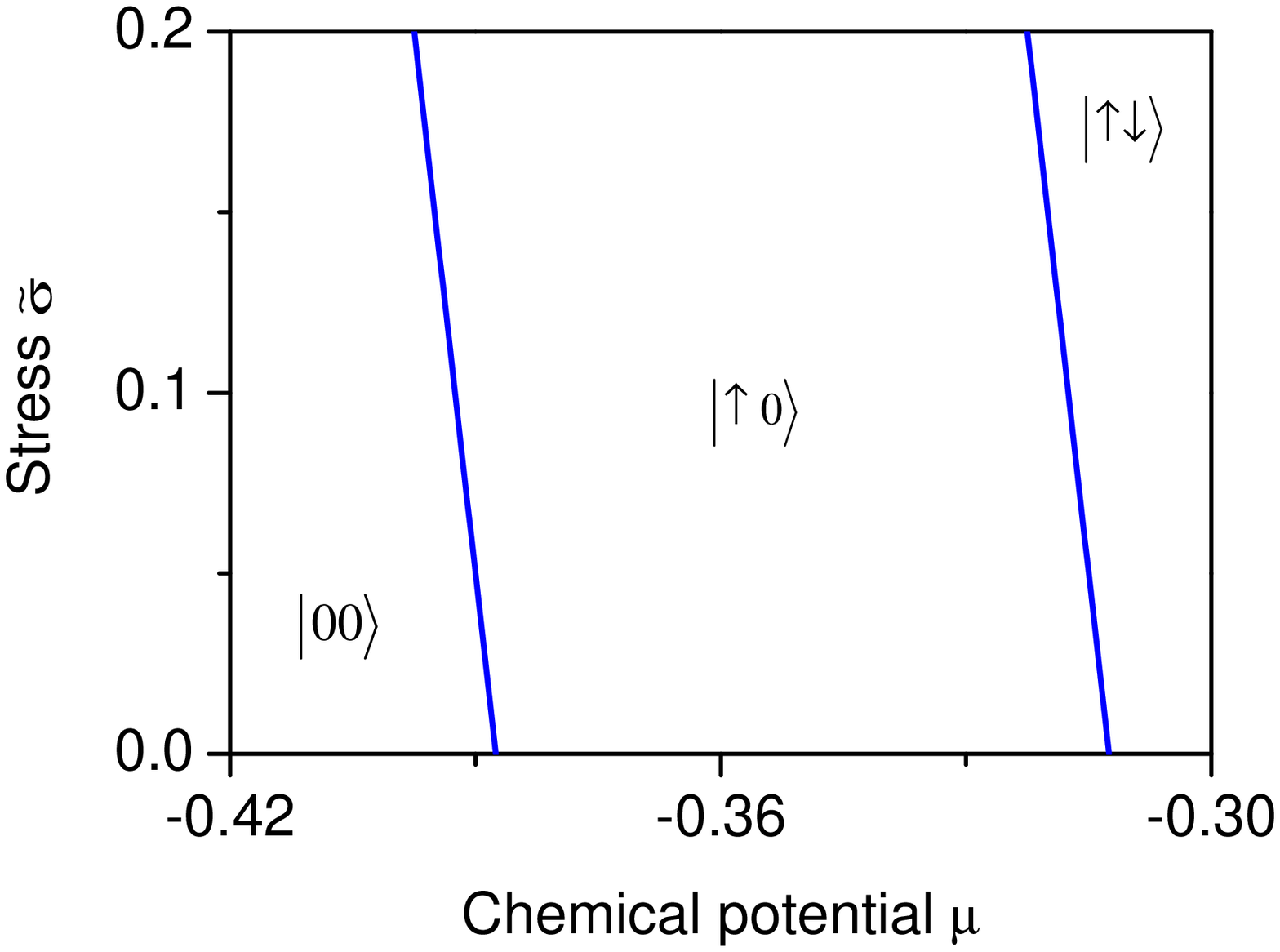}%
\hfill%
\includegraphics[width=0.47\textwidth]{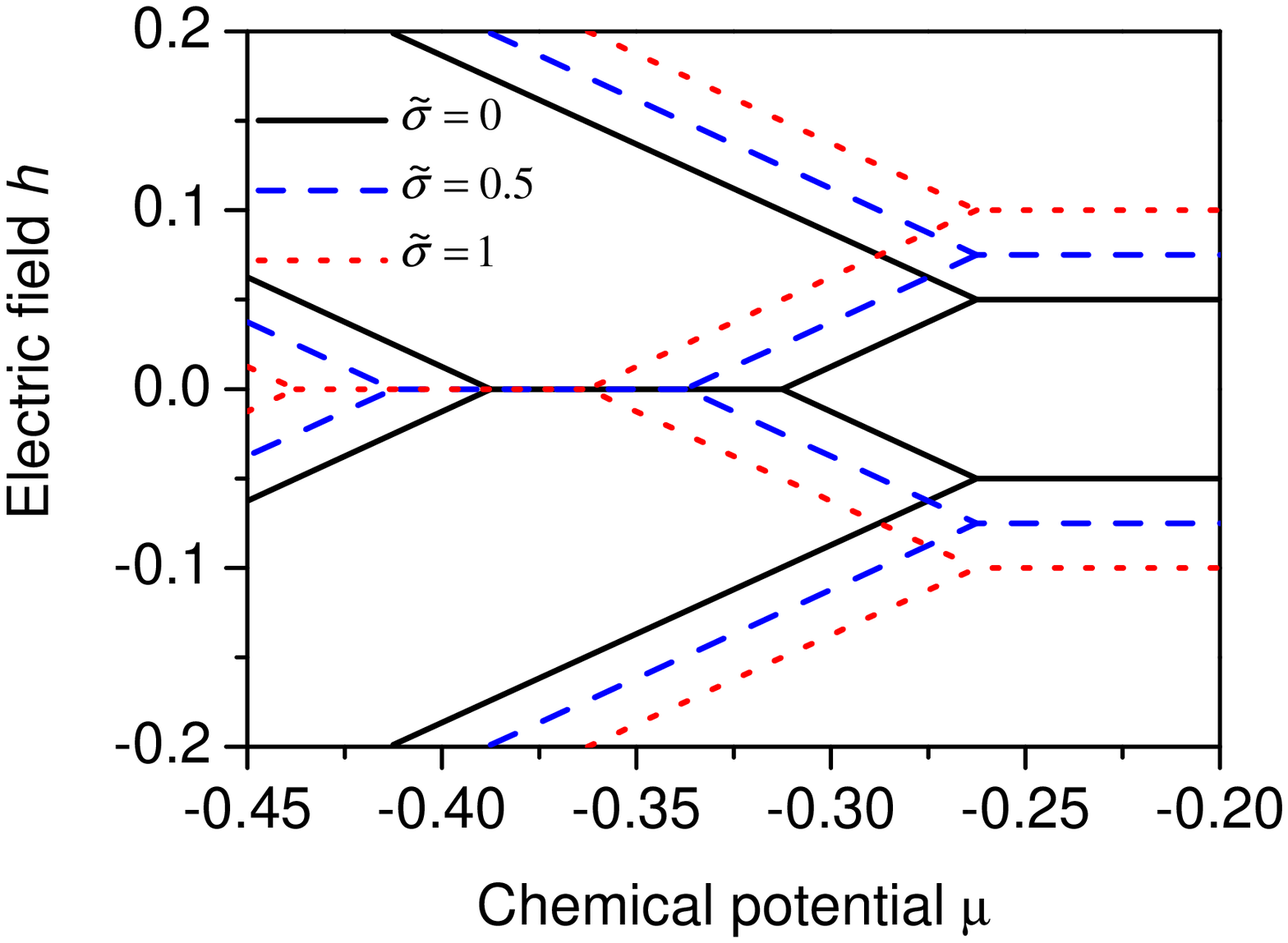}%
\caption{The effect of stress imposition: the ground state phase
diagram in the $\mu$--$\tilde{\sigma}$ coordinates at $h=0$ (left)
and a family of ground state phase diagrams for different values
of the stress (right). Other model parameters have the following
values: $W_{++}=1$, $W_{+-}=1.5$, $W_{--}=0.2$, $W_{-+}=0.3$,
$k_{\Delta}=0.05$.}
\label{fig:GSPD-MuSS}%
\end{figure}

\begin{wrapfigure}[17]{i}{0.5\textwidth}
\centerline{\includegraphics[width=0.47\textwidth]{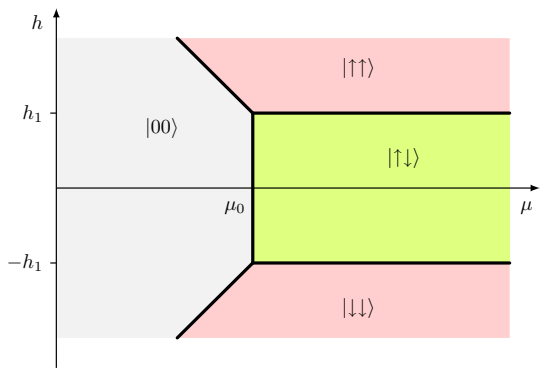}}%
\caption{A reduced form of the ground state phase diagram: only
the empty and full phases coexist.}
\label{fig:GSPD-B}%
\end{wrapfigure}

Setting equal thermodynamic potentials of different phases, one
can obtain a set of equations for the respective phase
transitions:
\begin{align}
    &\lvert 0 \rangle\leftrightarrow\lvert 1 \rangle\colon&
    \mu &\textstyle=-h -W_1 - k_{\Delta}(\tilde{\sigma}+\frac{1}{4}),&&&&
\notag\\%
    &\lvert 0 \rangle\leftrightarrow\lvert 3 \rangle\colon&
    \mu &\textstyle= h -W_1 - k_{\Delta}(\tilde{\sigma}+\frac{1}{4}),&&&&
\notag\\%
    &\lvert 0 \rangle\leftrightarrow\lvert 5 \rangle\colon&
    \mu &\textstyle=-h -W_2,&&&&
\notag\\%
    &\lvert 0 \rangle\leftrightarrow\lvert 6 \rangle\colon&
    \mu &\textstyle= h -W_2,&&&&
\notag\\%
    &\lvert 0 \rangle\leftrightarrow\lvert 7 \rangle\colon&
    \mu &\textstyle=-W_3 - k_{\Delta}(\tilde{\sigma}+\frac{1}{2}),&&&&
\notag\\%
    &\lvert 1 \rangle\leftrightarrow\lvert 3 \rangle\colon&
    h &=0,&&&&
\notag\\%
    &\lvert 5 \rangle\leftrightarrow\lvert 6 \rangle\colon&
    h &=0,&&&&
\notag\\%
    &\lvert 5 \rangle\leftrightarrow\lvert 7 \rangle\colon&
    \mu &\textstyle= (W_3-W_2) + k_{\Delta}(\tilde{\sigma}+\frac{1}{2}),&&&&
\notag\\%
    &\lvert 6 \rangle\leftrightarrow\lvert 7 \rangle\colon&
    \mu &\textstyle= -(W_3-W_2) - k_{\Delta}(\tilde{\sigma}+\frac{1}{2}),&&&&
\notag\\%
    &\lvert 1 \rangle\leftrightarrow\lvert 5 \rangle\colon&
    \mu &\textstyle=-h +(W_1-2W_2) + k_{\Delta}(\tilde{\sigma}+\frac{1}{4}),&&&&
\notag\\%
    &\lvert 3 \rangle\leftrightarrow\lvert 6 \rangle\colon&
    \mu &\textstyle=h +(W_1-2W_2) + k_{\Delta}(\tilde{\sigma}+\frac{1}{4}),&&&&
\notag\\%
    &\lvert 1 \rangle\leftrightarrow\lvert 7 \rangle\colon&
    \mu &\textstyle=h +(W_1-2W_3)  k_{\Delta}(\tilde{\sigma}+\frac{3}{4}),&&&&
\notag\\%
    &\lvert 3 \rangle\leftrightarrow\lvert 7 \rangle\colon&
    \mu &\textstyle=-h +(W_1-2W_3)  k_{\Delta}(\tilde{\sigma}+\frac{3}{4}),&&&&
\notag\\%
    &\lvert 1 \rangle\leftrightarrow\lvert 6 \rangle\colon&
    \mu &\textstyle=3h +(W_1-2W_2) + k_{\Delta}(\tilde{\sigma}+\frac{1}{4}),&&&&
\notag\\%
    &\lvert 3 \rangle\leftrightarrow\lvert 5 \rangle\colon&
    \mu &\textstyle=-3h +(W_1-2W_2) + k_{\Delta}(\tilde{\sigma}+\frac{1}{4}).&&&&
\label{eq:GSPTeqs}%
\end{align}
Various possible phase diagrams of the ground state in the
$\mu$--$h$ plane (which are calculated from equations
\eqref{eq:GSPTeqs}) are depicted in
figures~\ref{fig:GSPD-C}--\ref{fig:GSPD-B} using the following
designations:
\begin{align}
    \mu_0 &\textstyle= -W_3 - k_{\Delta}(\tilde{\sigma}+\frac{1}{2}),
\notag\\%
    \mu_1 &\textstyle= -W_1 - k_{\Delta}(\tilde{\sigma}+\frac{1}{4}),
\notag\\%
    \mu_2 &\textstyle= W_1 -2W_3 - k_{\Delta}(\tilde{\sigma}+\frac{3}{4}),
\notag\\%
    \mu_3 &\textstyle= W_1 -2W_2 + k_{\Delta}(\tilde{\sigma}+\frac{1}{4}),
\notag\\%
    h_1 &\textstyle= W_3 -W_2 + k_{\Delta}(\tilde{\sigma}+\frac{1}{2}).
\label{eq:GSPTfig}%
\end{align}
It is evident that the form of the phase diagram depends on the
values of model parameters, so we shall analyse some important
cases below.

In the most general case (figure~\ref{fig:GSPD-C}), the phase
diagram consists of empty, half-filled, and full states which
differ in occupation and polarization of sublattices.
Two-sublattice nature of the model demonstrates itself in the
intermediate half-filled states and the central nonpolar full
state thus being noticeably different from the respective ground
state diagram of the BEG model. It should be stressed that due to
the exclusion of the states 2, 4, and 8 the phase
$\lvert{\uparrow}0\rangle\equiv\lvert1\rangle$ is symmetrical to
the phase $\lvert0{\downarrow}\rangle\equiv\lvert3\rangle$. In the
considered case, both transitions between the empty and
half-filled phases as well as between the empty and full phases
are possible. The latter transition takes place, e.g.\ at zero
field $h$ which is contrary to the situation in the lithiated
anatase where only half-filled phases are accessible.

Coexistence of the Li-poor and Li-rich phases in the lithiated
anatase is fairly described by a phase diagram in
figure~\ref{fig:GSPD-A}. For any value of the external electric
field $h$ (in particular, at zero field) the system can pass from
the empty state to the half-filled state only. In the regime of
fixed concentration such a phase transition manifests itself as a
phase separation.

Such a phase diagram exists if conditions $\mu_1<\mu_2<\mu_3$ are
satisfied. Taking into account definitions \eqref{eq:GSPTfig} and
\eqref{eq:W123} and neglecting deformational effects one can
rewrite the above inequality in a more clear form
\begin{equation}
    W_{12}^{12} < W_{12}^{11} < 0.
\label{eq:GSPTWWcond1}%
\end{equation}
Thus, this type of a phase diagram exists if the interactions
between the orientational states in different sublattices are repulsive
with the interaction between the unlike states (with different
$p$, i.e.\ $\lvert{\uparrow}0\rangle$ and
$\lvert0{\uparrow}\rangle$ or $\lvert{\downarrow}0\rangle$ and
$\lvert0{\downarrow}\rangle$) being stronger than the interaction
between the like states (with identical $p$, i.e.\
$\lvert{\uparrow}0\rangle$ and $\lvert0{\downarrow}\rangle$ or
$\lvert{\downarrow}0\rangle$ and $\lvert0{\uparrow}\rangle$).
These conclusions are in full qualitative agreement with the
predictions of the molecular dynamics modelling
\cite{tdio:Wagemaker03}.

In view of deformational effects, the inequalities
$\mu_1<\mu_2<\mu_3$ transform to
\begin{equation}
    \textstyle
    -k_{\Delta}(\tilde{\sigma}+\frac{1}{4}) <
    -W_{12}^{11}-k_{\Delta}(\tilde{\sigma}+\frac{3}{4}) <
    -W_{12}^{12}+k_{\Delta}(\tilde{\sigma}+\frac{1}{4}).
\label{eq:GSPTWWcond2}%
\end{equation}
Thus, application of the stress $\tilde{\sigma}$ favours the phase
diagram in figure~\ref{fig:GSPD-A} because $\mu_1$ and $\mu_2$ are
shifted to the left $\sim \tilde{\sigma}$ and $\mu_3$ is moved
towards the right $\sim \tilde{\sigma}$. So the domains of the
phases $\lvert{\uparrow}0\rangle$ and $\lvert0{\downarrow}\rangle$
expand. However, at $h=0$ the difference $\mu_2-\mu_1 =
-W_{12}^{11}-\frac{1}{2}k_{\Delta}$ does not depend on the stress
(as one can see on the phase diagram $\mu$--$\tilde{\sigma}$ in
figure~\ref{fig:GSPD-MuSS}) and half-filled phases exist if
$W_{12}^{11}<-\frac{1}{2}k_{\Delta}$. A strong enough stress
$\tilde{\sigma}$ can ``open'' these phases even if conditions
\eqref{eq:GSPTWWcond1} are not satisfied. The above
rationales are confirmed by the family of phase diagrams in
figure~\ref{fig:GSPD-MuSS} calculated for different stress values
$\tilde{\sigma}$ (hereinafter all model parameters are given in
the dimensionless units normalized by $W_{++}=1$; in the
considered case $W_{11}^{11}=0.75$, $W_{11}^{12}=0.5$,
$W_{12}^{11}=-0.1$, $W_{12}^{12}=-0.15$ in such a manner
satisfying conditions \eqref{eq:GSPTWWcond1}).

If conditions \eqref{eq:GSPTWWcond2} fail, the half-filled phases
are suppressed (figure~\ref{fig:GSPD-B}). In addition, if $h_1
\leqslant 0$, then the central full nonpolar phase vanishes and
the ground state phase diagram reduces to the respective one for
the BEG model.
\vspace{7ex}

\subsection{Temperature behaviour of phase diagrams and phase separation}

\begin{figure}
\centerline{\includegraphics[width=0.8\textwidth]{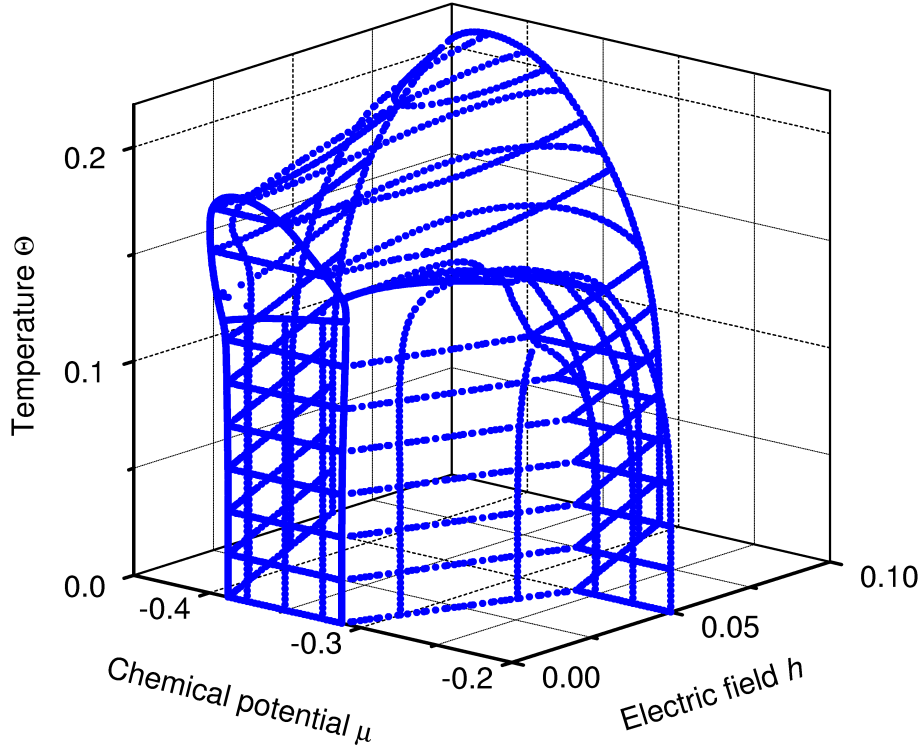}}%
\caption{The three-dimensional phase diagram in the
$\mu$--$h$--$\Theta$ coordinates. Model parameters have the
following values: $W_{++}=1$, $W_{+-}=1.5$, $W_{--}=0.2$,
$W_{-+}=0.3$, $k_{\Delta}=0.05$, $\tilde{\sigma}=0$.}%
\label{fig:PhD-MuhT}%
\end{figure}

Since the considered model inherits features of the Mitsui and BEG
models, one can expect a quite complex thermodynamical behaviour
and the above analysis of the ground state has proved these
anticipations. As it is seen from the temperature axis
complemented analogue (figure~\ref{fig:PhD-MuhT}) of the diagram
in figure~\ref{fig:GSPD-A}, the lines of phase transitions form
the surfaces and some new phases appear. A comprehensive analysis
of the obtained diagram is too complicated and goes beyond the
scope of this research, so we consider the case with intermediate
half-filled phases corresponding to the intercalated anatase.

\begin{wrapfigure}[25]{i}{0.5\textwidth}
\vspace*{3ex}
\centerline{\includegraphics[width=0.49\textwidth]{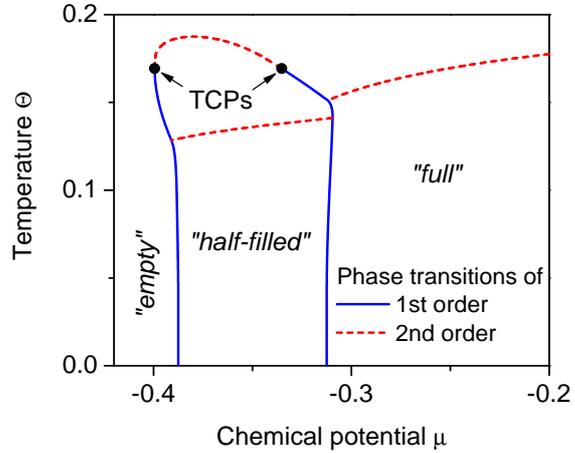}}%
\vspace{3ex}
\caption{The phase diagram $\mu$--$\Theta$ at the absence of the
electric field. Other model parameters are as follows: $W_{++}=1$,
$W_{+-}=1.5$, $W_{--}=0.2$, $W_{-+}=0.3$, $k_{\Delta}=0.05$,
$\tilde{\sigma}=0$. Phase names (``empty'', ``half-filled'', and
``full'') correspond to the ground state. There are two
tricritical points (TCPs) on the phase transition line limiting
the ``half-filled''
phase from the top.}%
\label{fig:PhD-zeroh}%
\end{wrapfigure}

With this in mind let us analyse the phase diagram at the absence
of the external electric field $h$ (figure~\ref{fig:PhD-zeroh}).
At low temperature, in full agreement with the ground state
diagram, there are three phases: ``empty'', ``half-filled'', and
``full'' (due to the temperature ``blurring'' at higher
temperatures these names loose their exact meaning). At high
temperatures in the whole range of chemical potential there is
only ``empty'' phase where all order parameters are zero except
the concentration $n_{+}$ which changes monotonously.

\clearpage

\begin{figure}
\includegraphics[width=0.47\textwidth]{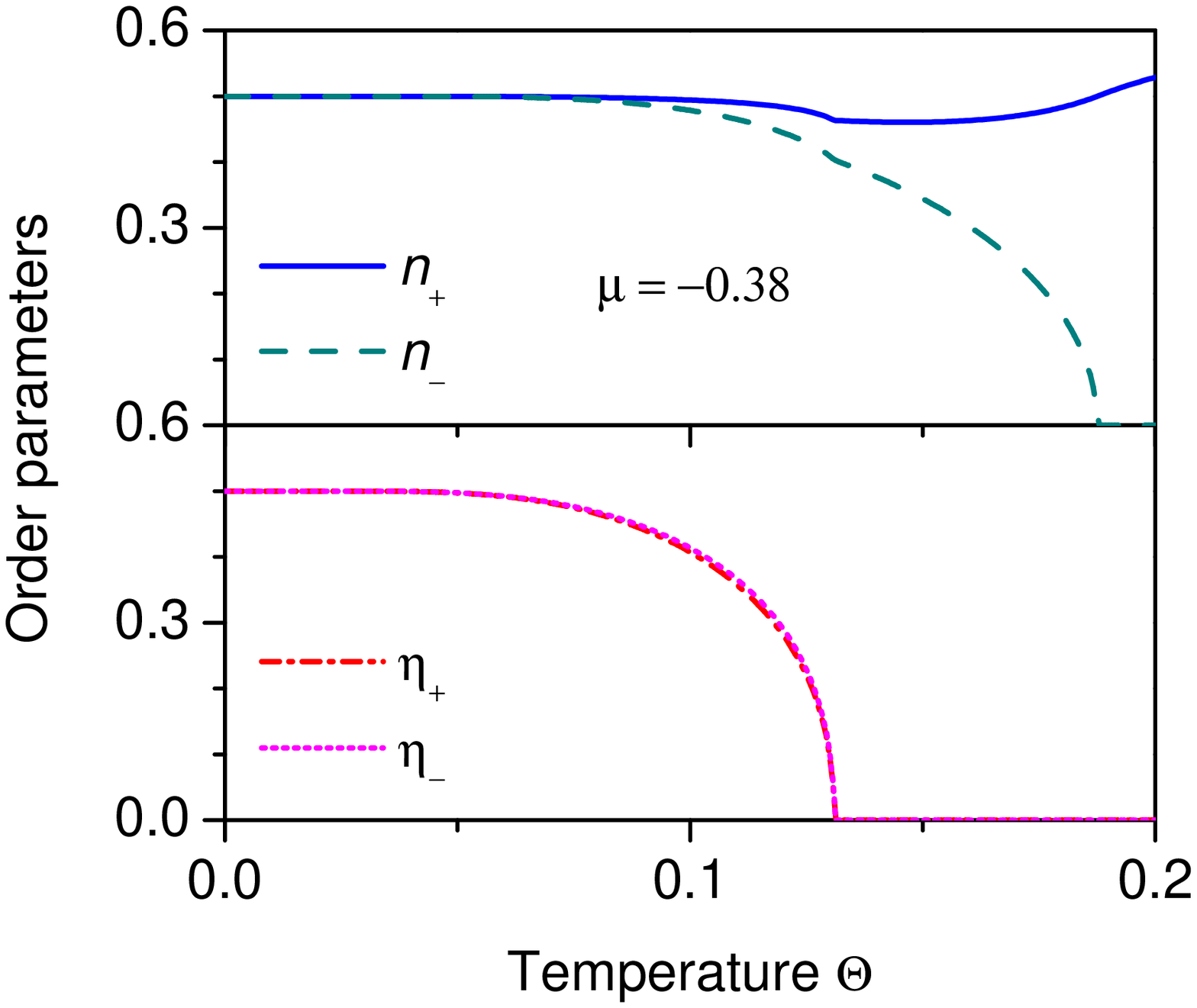}%
\hfill%
\includegraphics[width=0.48\textwidth]{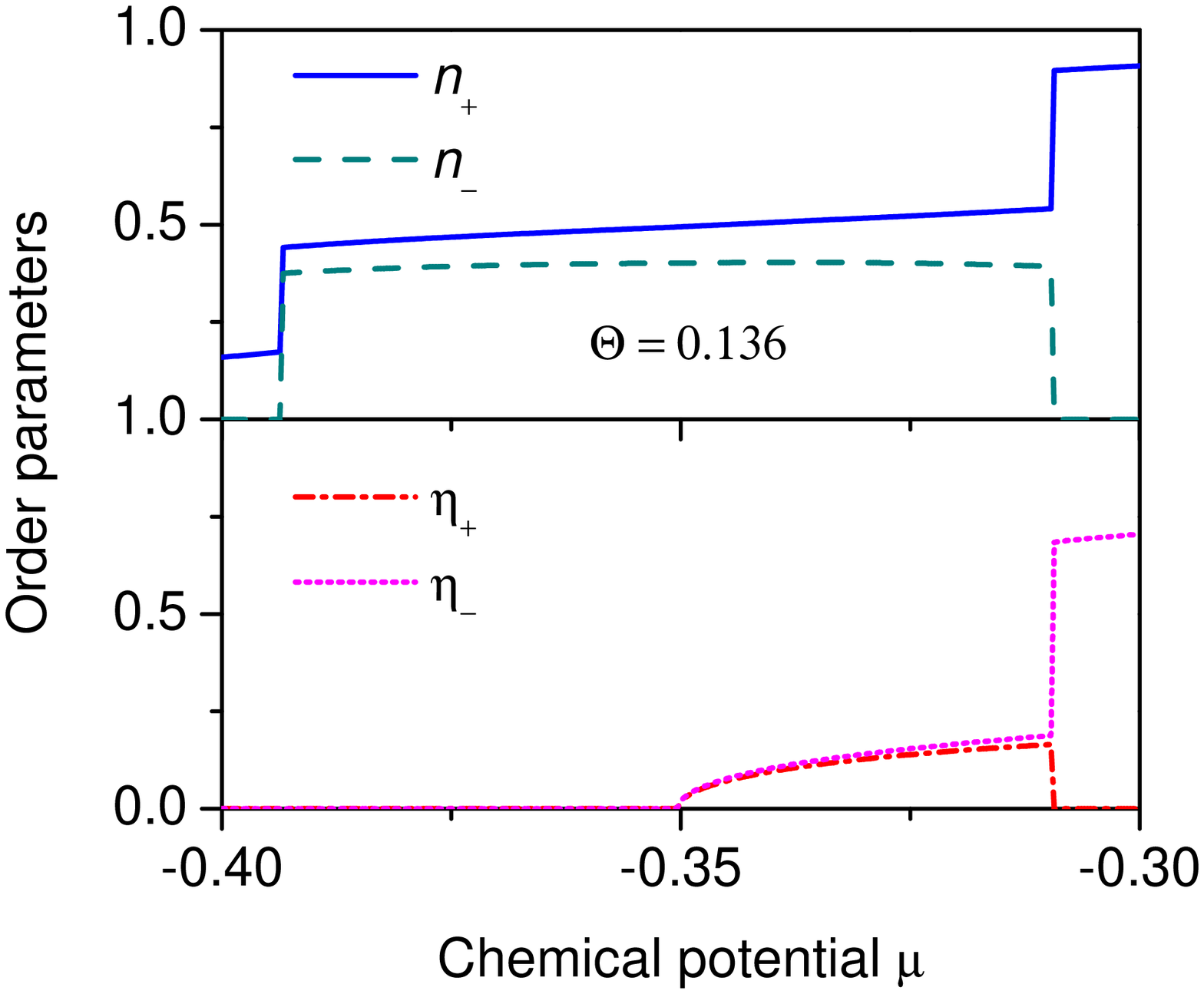}%
\caption{Phase diagram ``cross-sections'': temperature dependences
of order parameters in the ``half-filled'' phase (left:
$\mu=-0.38$; in the picture scale curves $\eta_{+}$ and $\eta_{-}$
overlap) and dependences of order parameters on chemical potential
exhibiting phase transitions of the second and the first orders
(right: $\Theta=0.136$). Other model parameters have the following
values: $W_{++}=1$, $W_{+-}=1.5$, $W_{--}=0.2$, $W_{-+}=0.3$,
$k_{\Delta}=0.05$, $\tilde{\sigma}=0$.}%
\label{fig:Sctn}%
\end{figure}

The ``half-filled'' phase provides the most complex behaviour of
the order parameters. As is seen from the ``cross-section'' at the
chemical potential value $\mu=-0.38$ (figure~\ref{fig:Sctn}), the
rise of temperature leads to the suppression of the dipole-dipole
ordering (order parameters $\eta_{+}$ and $\eta_{-}$) and it
completely vanishes at the line of the second order phase
transition which is located inside the ``half-filled'' phase. The
phase itself is limited from the top side by the line of the phase
transition with zeroing of $n_{-}$. The upper part of the phase is
separated by the tricritical points (TCPs) marking the change of
the phase transition order from the second to the first one. The
``full'' phase is also limited from the top side by the line of
the second order transitions where  $\eta_{-}\to0$.

Behaviour of order parameters at the change of chemical potential
(figure~\ref{fig:Sctn}) clearly distinguishes the phases separated
by the lines of the first order phase transitions (e.g.\ a
characteristic feature of the ``half-filled'' phase is
$n_{-}\neq0$). In a certain temperature range inside the
``half-filled'' phase, the above mentioned phase transition
between the polar and non-polar states takes place.

\begin{figure}[!b]
\includegraphics[width=0.46\textwidth]{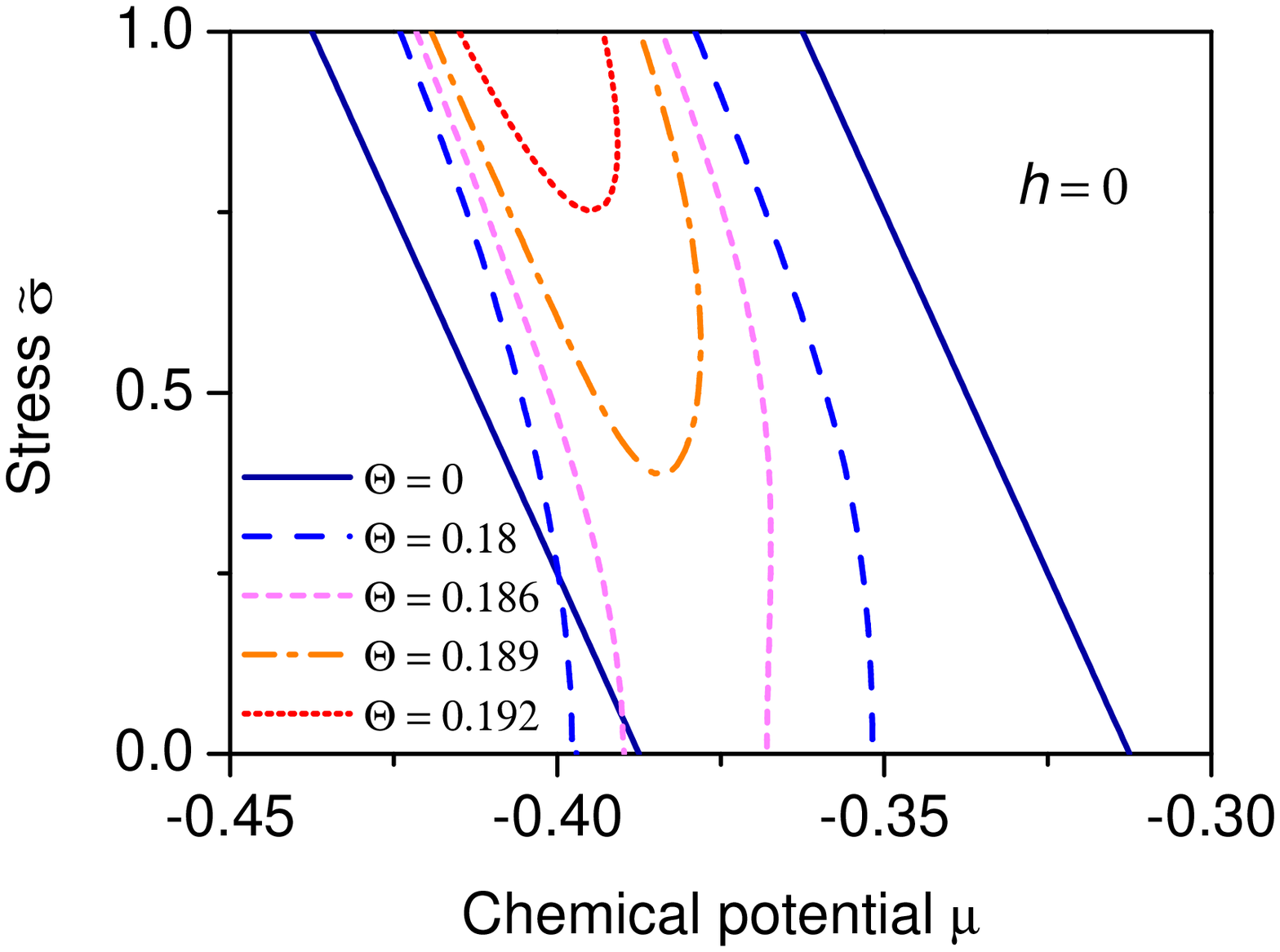}%
\hfill%
\includegraphics[width=0.48\textwidth]{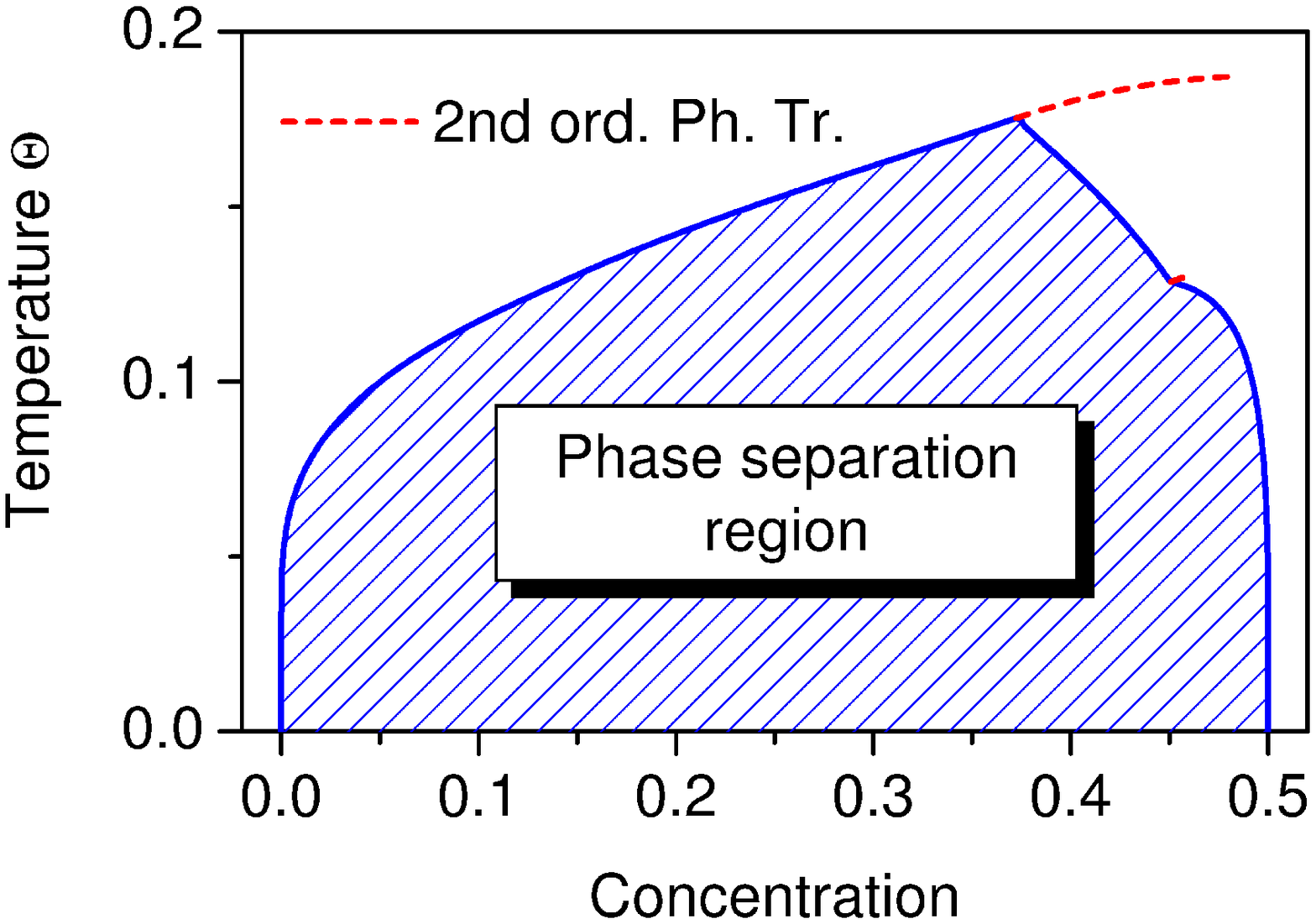}%
\\%
\parbox[t]{0.48\textwidth}{%
\caption{A family of phase diagrams ``chemical potential $\mu$ --
stress $\tilde{\sigma}$'' for various temperature values at $h=0$.
Other model parameters have the following values: $W_{++}=1$,
$W_{+-}=1.5$, $W_{--}=0.2$,
$W_{-+}=0.3$, $k_{\Delta}=0.05$.}%
\label{fig:Mu-S-zh}%
}%
\hfill%
\parbox[t]{0.47\textwidth}{%
\caption{The diagram of the phase separation into ``poor''
($n_{+}=0$) and ``rich'' ($n_{+}=0.5$) phases in the regime
$n_{+}=\mathrm{const}$ ($W_{++}=1$, $W_{+-}=1.5$, $W_{--}=0.2$,
$W_{-+}=0.3$, $k_{\Delta}=0.05$, $\tilde{\sigma}=0$).}%
\label{fig:PhSep}%
}
\end{figure}

In a wide temperature range, the appearance of the phase diagram
``chemical potential $\mu$ -- stress $\tilde{\sigma}$''
(figure~\ref{fig:Mu-S-zh}) closely resemble the ground state one
(figure~\ref{fig:GSPD-MuSS}). However, the further rise of
temperature leads to a fast suppression of the ``half-filled''
phase.

Since in the regime of a fixed chemical potential
($\mu=\mathrm{const}$) the ``empty'' and ``half-filled'' phases on
the phase diagram $\mu$--$\Theta$ (figure~\ref{fig:PhD-zeroh}) are
separated mainly by the line of the first order phase transitions,
the system separates into ``poor'' and ``rich'' phases
(figure~\ref{fig:PhSep}) in the regime of the fixed concentration
($n_{+}=\mathrm{const}$). As one can see, in a wide region of low
temperatures a separation into concentrations $n_{+}=0$ and
$n_{+}=0.5$ occurs which well reproduces the coexistence of
Li-poor and Li-rich phases in the intercalated anatase. The phase
separation region narrows at heating and finally closes up at the
point corresponding to the tricritical point in
figure~\ref{fig:PhD-zeroh}. Another short line of phase
transitions (starting at the kink of the right boundary) relates
to the suppression of polar states in a ``half-filled'' phase.

\clearpage

\begin{wrapfigure}{i}{0.5\textwidth}
\centerline{\includegraphics[width=0.47\textwidth]{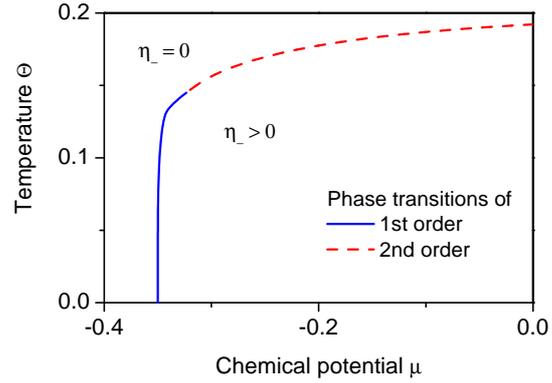}}%
\caption{The phase diagram $\mu$-$\Theta$ at $h=0$ for the case of
the first order phase transition between ``empty'' and ``full''
phases ($n_{-}=0$ and $\eta_{+}=0$ in both of them; $W_{++}=1$,
$W_{+-}=0$, $W_{--}=0$, $W_{-+}=0.3$, $k_{\Delta}=0.05$,
$\tilde{\sigma}=0$). }%
\label{fig:PhD2-zeroh}%
\end{wrapfigure}
\begin{figure}[h]
\includegraphics[width=0.47\textwidth]{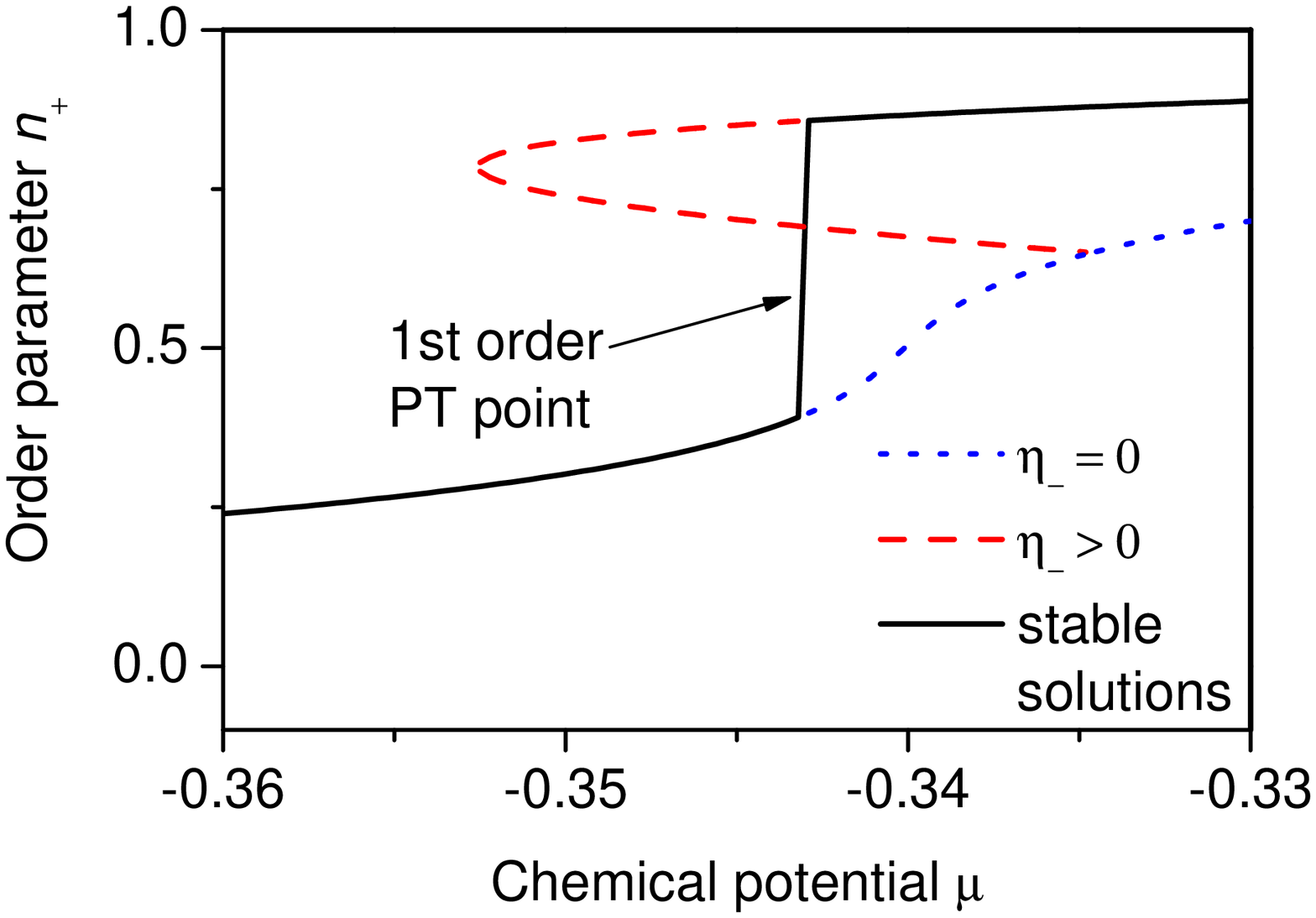}%
\hfill%
\includegraphics[width=0.47\textwidth]{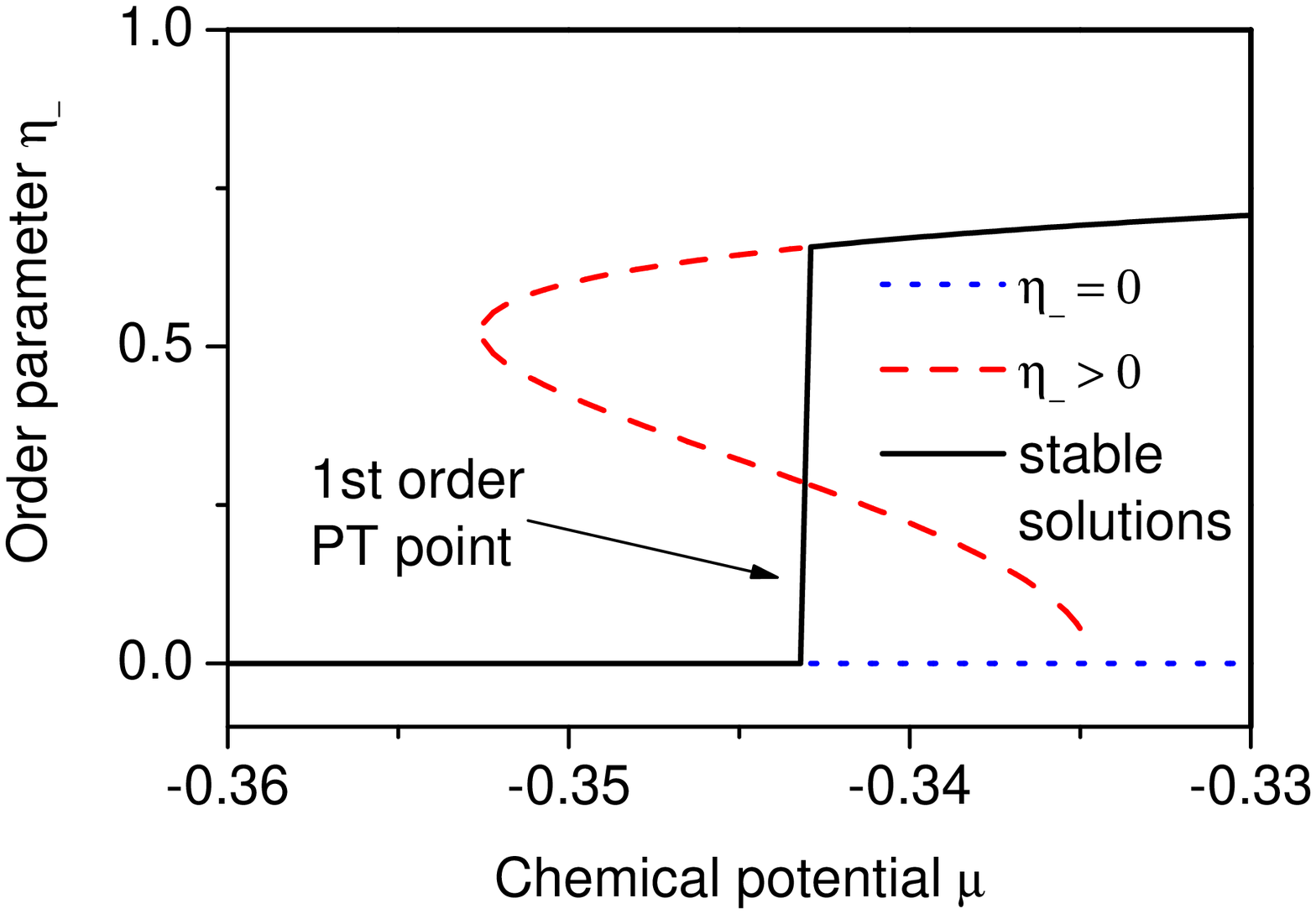}%
\caption{A jump of order parameters $n_{+}$ (left) and $\eta_{-}$
(right) at the first order phase transition illustrating the
previous phase diagram. Thermodynamically stable solutions are
marked with the solid curve. Other model parameters have the
following values: $W_{++}=1$, $W_{+-}=0$, $W_{--}=0$,
$W_{-+}=0.3$, $k_{\Delta}=0.05$, $\Theta=0.13$, $h=0$,
$\tilde{\sigma}=0$.}
\label{fig:RhoEta2}
\end{figure}

A question arises why in the microscopic model we should deal with
four order parameters while the Landau expansion is quite
successful with only two? First of all, the semiphenomenological
description is qualitative only and it just demonstrates the
possibility of the first order phase transition with simultaneous
jumps of concentration and the order parameter $\eta$ as the
minimal set necessary to describe the phase separation in the
litiated anatase. Such a picture corresponds to a direct phase
transition between the ``empty'' and ``full'' phases (see the
phase diagram at zero external field in
figure~\ref{fig:PhD-zeroh}) when other order parameters ($n_{-}$
and $\eta_{+}$) are always equal to zero. The dependences of the
``active'' order parameters $n_{+}$ and $\eta_{-}$ on chemical
potential (figure~\ref{fig:RhoEta2}) closely resemble the
respective curves for $\rho$ and $\eta$ obtained by the Landau
expansion (figure~\ref{fig:RhoEta}). But if one should take into
account the ``half-filled'' phase (what is inevitable for
description of the lithiated anatase), the order parameters
($n_{-}$ and $\eta_{+}$) became nonzero and the full set of four
parameters should be considered as it has been done above.

\section{Conclusions}

To sum up, the present study was inspired by two features of the
lithium intercalated anatase: coexistence of poor and rich phases
and two possible localizations of Li ion in the oxygen octahedron
along the $c$ axis. The second one implies the possibility of
(anti)polar ordering what is beyond the scope of a simple lattice
gas model well describing a phase separation. So a model of
Blume-Emery-Griffits type has been used which takes into account
both particle-particle and dipole-dipole terms.

Another less obvious peculiarity of the considered compound is the
intercalation induced deformation of lattice: Li-rich phase has a
lower symmetry (the axes $a$ and $b$ become inequivalent)
resulting in preferential occupation of one position of the
mentioned pairs which has an alternating orientation in the
neighbour octahedra (i.e.\ in different sublattices). Performed
symmetry analysis explains this phenomena by the possibility of
internal piezoeffect: the deformation in the $ab$ plane as well as
the appearance of an effective internal staggered field (causing
the ordering of antiferroelectric type like as in the Mitsui
model) both belong to the same irreducible representation of the
initial high-symmetry anatase phase and, hence, are described by
the common order parameter. Thus, increase of the intercalant
content could result in a phase transition with simultaneous jumps
of the average occupation and antipolarization (the latter
accompanied by the jump of dielectric susceptibility) as it has
been proved by the Landau expansion.

The microscopic approach, combining the abovementioned features of
both the BEG and Mitsui models, gives semiquantitative description
of phase coexistence in the lithiated anatase. Analysis of the
ground state phase diagram confirms a possibility of the phase
transition between ``empty'' and ``half-filled'' phases which
corresponds to the phase separation into Li-poor and Li-rich
phases in the crystal. As the model predicts, such a separation
remains near constant in a wide temperature range. The microscopic
approach could easily reproduce the Landau expansion results as
the particular case of the ``empty''-``full'' transition described
by the two order parameters. But the presence of the
``half-filled'' phase makes it necessary to deal with the full set
of the order parameters allowed by the crystal symmetry.

However, some issues are still open. The model predicts that
external stress should shift the phase transition between empty
and half-filled phases to the lower values of chemical potential.
An experimental evidence of this conjecture is still missing. The
real average occupation in the Li-rich phase is 0.55--0.6 instead
of the value $n_{+}=0.5$ in the half-filled model phase. This
deviation could be explained by a multidomain nature of the
Li-rich phase containing ``impurities'' of the full-occupied
LiTiO$_2$ phase while the model phases are monodomain by
definition. The same explanation applies to the issue of absence
of the total polarization in the half-filled phases: at zero
external field these phases with opposite polarizations could
coexist in different domains providing a full mutual compensation.
A similar mechanism of the mutual compensation of polarization in
Li-``chains'' with an opposite Li orientation is supported by the
molecular dynamics simulations \cite{tdio:Wagemaker03}.

\section*{Acknowledgements}

Oleh Velychko is deeply indebted to the project ``Improvement of
functional possibilities of the Western coordinating grid-centre
of Ukrainian Academic Grid (the City of Lviv)'' for financial
support and a possibility to perform numerical calculations on its
cluster.

\clearpage

\appendix

\section*{\mathversion{bold}%
Appendix:
Table of the point group $D_{4h}$ symmetry transformations for the
symmetrized averages}
\label{sect:SymOp}

\begin{table}[!h]
\caption{Transformations of the symmetrized linear combinations of
the averages which correspond to the irreducible representations
(IR) of the point symmetry group $D_{4h}$. The operations, which
also belong to the lower symmetry subgroup $D_{2h}$, are marked by
asterisk; they keep the symmetrized combination $\eta_{-}$ (IR
${B}_{1g}$) invariant.}
\label{tab:SymOp}
\bigskip%
\begin{center}
\begin{tabular}{|r||r|r|r|r|r|r|r|r|}
    \hline
    IR          & $\mathstrut^{\ast}$$E$    & $\mathstrut^{\ast}$$C_2^{(z)}$ & $C_4$       & $C_4^3$     & $\mathstrut^{\ast}$$C_2^{(y)}$ & $\mathstrut^{\ast}$$C_2^{(x)}$ & $C_2^{(xy)}$ & $C_2^{(x\bar{y})}$
\\
    \hline
    \hline
    ${A}_{1g}$  & $n_{+}$                   & $n_{+}$                        & $n_{+}$     & $n_{+}$     & $n_{+}$                        & $n_{+}$                        & $n_{+}$      & $n_{+}$
\\
    ${B}_{2u}$  & $n_{-}$                   & $n_{-}$                        & $-n_{-}$    & $-n_{-}$    & $-n_{-}$                       & $-n_{-}$                       & $n_{-}$      & $n_{-}$
\\
    ${A}_{2u}$  & $\eta_{+}$                & $\eta_{+}$                     & $\eta_{+}$  & $\eta_{+}$  & $-\eta_{+}$                    & $-\eta_{+}$                    & $-\eta_{+}$  & $-\eta_{+}$
\\
    ${B}_{1g}$  & $\eta_{-}$                & $\eta_{-}$                     & $-\eta_{-}$ & $-\eta_{-}$ & $\eta_{-}$                     & $\eta_{-}$                     & $-\eta_{-}$  & $-\eta_{-}$
\\
    \hline
\end{tabular}
\end{center}
\begin{center}
\begin{tabular}{|r||r|r|r|r|r|r|r|r|}
    \hline
    IR          & $\mathstrut^{\ast}$$I$ & $\mathstrut^{\ast}$$m_{(xy)}$ & $S_4^3$     & $S_4$       & $\mathstrut^{\ast}$$m_{(xz)}$ & $\mathstrut^{\ast}$$m_{(yz)}$ & $m_{(x\bar{y})}$ & $m_{(xy)}$
\\
    \hline
    \hline
    ${A}_{1g}$  & $n_{+}$                & $n_{+}$                       & $n_{+}$     & $n_{+}$     & $n_{+}$                       & $n_{+}$                       & $n_{+}$          & $n_{+}$
\\
    ${B}_{2u}$  & $-n_{-}$               & $-n_{-}$                      & $n_{-}$     & $n_{-}$     & $n_{-}$                       & $n_{-}$                       & $-n_{-}$         & $-n_{-}$
\\
    ${A}_{2u}$  & $-\eta_{+}$            & $-\eta_{+}$                   & $-\eta_{+}$ & $-\eta_{+}$ & $\eta_{+}$                    & $\eta_{+}$                    & $\eta_{+}$       & $\eta_{+}$
\\
    ${B}_{1g}$  & $\eta_{-}$             & $\eta_{-}$                    & $-\eta_{-}$ & $-\eta_{-}$ & $\eta_{-}$                    & $\eta_{-}$                    & $-\eta_{-}$      & $-\eta_{-}$
\\
    \hline
\end{tabular}
\end{center}
\end{table}

\bibliographystyle{cmp01}
\bibliography{tio2,review}

\label{last@page}

\end{document}